\documentclass[journal,twoside]{IEEEtran}

\usepackage{mathrsfs}
\usepackage[noadjust]{cite}
\usepackage{graphicx,color,overpic,psfrag}
\usepackage{amsmath, amssymb}
\usepackage{latexsym}
\usepackage{bm}
\usepackage{amssymb}
\usepackage{cases}
\usepackage{array}
\usepackage{fancyhdr}
\usepackage{setspace}

\ifCLASSOPTIONcompsoc
\usepackage[caption=false,font=normalsize,labelfont=sf,textfont=sf]{subfig}
\else
\usepackage[caption=false,font=footnotesize]{subfig}
\fi

\usepackage{url}
\usepackage{algpseudocode}
\usepackage{algorithm}
\usepackage{blkarray}
\usepackage{booktabs}

\usepackage{multirow}
\usepackage{dsfont}
\usepackage{tabularx}
\usepackage[table]{xcolor}

\usepackage{amsfonts}

\usepackage{letltxmacro}

\graphicspath{{figure/}}

\newcommand{\ra}[1]{\renewcommand{\arraystretch}{#1}}

\newtheorem{prop}{Proposition}

\newcommand{\figref}[1]{Fig. \ref{#1}}
\newcommand{\tabref}[1]{Table \ref{#1}}
\newcommand{\alref}[1]{Algorithm \ref{#1}}
\newcommand{\appref}[1]{Appendix \ref{#1}}
\newcommand{\secref}[1]{Section \ref{#1}}

\newcommand{\propref}[1]{Proposition \ref{#1}}

\newcommand{\Exp}{{\mathsf{E}}}
\newcommand{\expect}[1]{\Exp\left\{#1\right\}}

\newcommand{\diag}[1]{\mathsf{diag}\left\{#1\right\}}

\newcommand{\abs}[1]{\left|#1\right|}
\newcommand{\sqrabs}[1]{\left|#1\right|^{2}}

\newcommand{\oneon}[1]{\frac{1}{#1}}
\newcommand{\zeroldots}[1]{0,1,\ldots,#1}
\newcommand{\intd}{\mathrm{d}}
\newcommand{\intdx}[1]{\intd #1}

\newcommand{\GCN}[2]{\mathcal{CN}\left( #1 , #2\right) }

\newcommand{\argmin}[1]{\mathop{\arg\min}\limits_{#1}}

\newcommand{\ith}[1]{$#1$th}

\newcommand{\toinf}[1]{#1 \to\infty}

\newcommand{\delfunc}[1]{\delta\left(#1\right)}
\newcommand{\vecele}[2]{\left[#1\right]_{#2}}
\newcommand{\vecop}[1]{\mathsf{vec}\left\{#1\right\}}

\newcommand{\vecherop}[1]{\mathsf{vec}^{H}\left\{#1\right\}}

\newcommand{\modulo}[2]{\left\langle#1\right\rangle_{#2}}

\newcommand{\setdif}[2]{#1\backslash #2}

\newcommand{\kmh}[1]{#1 km/h}

\newcommand{\equaa}{\mathop{=}^{(\mathrm{a})}}
\newcommand{\equab}{\mathop{=}^{(\mathrm{b})}}
\newcommand{\equac}{\mathop{=}^{(\mathrm{c})}}
\newcommand{\equad}{\mathop{=}^{(\mathrm{d})}}
\newcommand{\equae}{\mathop{=}^{(\mathrm{e})}}

\newcommand{\cA}{\mathcal{A}}

\newcommand{\cK}{\mathcal{K}}

\newcommand{\ba}{\mathbf{a}}

\newcommand{\bbf}{\mathbf{f}}
\newcommand{\bg}{\mathbf{g}}

\newcommand{\bv}{\mathbf{v}}

\newcommand{\bx}{\mathbf{x}}

\newcommand{\bA}{\mathbf{A}}
\newcommand{\bB}{\mathbf{B}}
\newcommand{\bC}{\mathbf{C}}
\newcommand{\bD}{\mathbf{D}}

\newcommand{\bF}{\mathbf{F}}
\newcommand{\bG}{\mathbf{G}}
\newcommand{\bH}{\mathbf{H}}
\newcommand{\bI}{\mathbf{I}}

\newcommand{\bR}{\mathbf{R}}

\newcommand{\bU}{\mathbf{U}}
\newcommand{\bV}{\mathbf{V}}

\newcommand{\bX}{\mathbf{X}}
\newcommand{\bY}{\mathbf{Y}}
\newcommand{\bZ}{\mathbf{Z}}

\newcommand{\bbC}{\mathbb{C}}
\newcommand{\bbR}{\mathbb{R}}

\newcommand{\ttS}{\mathtt{S}}

\newcommand{\hatbG}{\hat{\bG}}
\newcommand{\hatbH}{\hat{\bH}}

\newcommand{\tildebH}{\tilde{\bH}}

\newcommand{\barbH}{\bar{\bH}}
\newcommand{\barbOmega}{\bar{\bOmega}}

\newcommand{\bzero}{\mathbf{0}}

\newcommand{\bOmega}{{\boldsymbol\Omega}}

\newcommand{\bPi}{{\boldsymbol\Pi}}

\newcommand{\fortext}{\textrm{for}}

\newcommand{\vtheta}{\bv_{M,\theta}}

\newcommand{\sintheta}{\sin\left(\theta\right)}

\newcommand{\expx}[1]{\exp\left(#1\right)}

\newcommand{\upnot}[2]{#1^{\mathrm{#2}}}
\newcommand{\dnnot}[2]{#1_{\mathrm{#2}}}

\newcommand{\barjmath}{\bar{\jmath}}

\newcommand{\ntb}{\notag\\}

\newcommand{\tabincell}[2]{\begin{tabular}{@{}#1@{}}#2\end{tabular}}

\newcommand{\singbrac}[1]{\left({#1}\right)}

\allowdisplaybreaks

\newcommand{\equadispwid}{5.5pt}

\begin{document}

\title{Channel Acquisition for Massive MIMO-OFDM with Adjustable Phase Shift Pilots}

\author{
Li~You,~\IEEEmembership{Student~Member,~IEEE,} Xiqi~Gao,~\IEEEmembership{Fellow,~IEEE,}
A.~Lee~Swindlehurst,~\IEEEmembership{Fellow,~IEEE,} and~Wen~Zhong%

\thanks{Copyright \copyright\  2015 IEEE. Personal use of this material is permitted. However, permission to use this material for any other purposes must be obtained from the IEEE by sending a request to pubs-permissions@ieee.org.}%
\thanks{This work was supported in part by the National Natural Science Foundation of China under Grants 61471113, 61320106003, and 61201171, the China High-Tech 863 Plan under Grants 2015AA01A701 and 2014AA01A704, the National Science and Technology Major Project of China under Grant 2014ZX03003006-003, and the Program for Jiangsu Innovation Team. The work of L. You was supported in part by the China Scholarship Council (CSC). This work was presented in part at the IEEE 16th International Workshop on Signal Processing Advances in Wireless Communications (SPAWC), Stockholm, Sweden, 2015.}%
\thanks{
L. You, X. Q. Gao, and W. Zhong are with the National Mobile Communications Research Laboratory, Southeast University, Nanjing 210096, China (e-mail: liyou@seu.edu.cn; xqgao@seu.edu.cn; wzhong@seu.edu.cn).
}
\thanks{
A. L. Swindlehurst is with the Center for Pervasive Communications and Computing (CPCC), University of California, Irvine, CA 92697 USA (e-mail: swindle@uci.edu).
}
}

\markboth{IEEE Transactions on Signal Processing}{YOU \MakeLowercase{\textit{et al.}}: Channel Acquisition for Massive MIMO-OFDM with Adjustable Phase Shift Pilots}

\maketitle

\begin{abstract}
We propose adjustable phase shift pilots (APSPs) for channel acquisition in wideband massive multiple-input multiple-output (MIMO) systems employing orthogonal frequency division multiplexing (OFDM) to reduce the pilot overhead. Based on a physically motivated channel model, we first establish a relationship between channel space-frequency correlations and the channel power angle-delay spectrum in the massive antenna array regime, which reveals the channel sparsity in massive MIMO-OFDM. With this channel model, we then investigate channel acquisition, including channel estimation and channel prediction, for massive MIMO-OFDM with APSPs. We show that channel acquisition performance in terms of sum mean square error can be minimized if the user terminals' channel power distributions in the angle-delay domain can be made non-overlapping with proper phase shift scheduling. A simplified pilot phase shift scheduling algorithm is developed based on this optimal channel acquisition condition. The performance of APSPs is investigated for both one symbol and multiple symbol data models. Simulations demonstrate that the proposed APSP approach can provide substantial performance gains in terms of achievable spectral efficiency over the conventional phase shift orthogonal pilot approach in typical mobility scenarios.
\end{abstract}


%
%
\section{Introduction}

\IEEEPARstart{F}{orthcoming} 5G cellular wireless systems are expected to support 1000 times faster data rates than the currently deployed 4G long-term evolution (LTE) system. To achieve the high data rates required by 5G, many technologies have been proposed \cite{Andrews14What,Boccardi14Five,Wang14Cellular}. Among them, massive multiple-input multiple-output (MIMO) systems, which deploy unprecedented numbers of antennas at the base stations (BSs) to simultaneously serve a relatively large number of user terminals (UTs), are believed to be one of the key candidate technologies for 5G \cite{Marzetta10Noncooperative,Larsson14Massive,Lu14overview}.

Orthogonal frequency division multiplexing (OFDM) is a multi-carrier modulation technology suited for high data rate wideband wireless transmission \cite{Cimini85Analysis,Stuber04Broadband}. Due to its robustness to channel frequency selectivity and relatively efficient implementation, OFDM combined with massive MIMO is a promising technique for wideband massive MIMO transmission \cite{Marzetta10Noncooperative}. As in conventional MIMO-OFDM, the performance of massive MIMO-OFDM is highly dependant on the quality of the channel acquisition. Pilot design and channel acquisition for massive MIMO-OFDM is of great practical importance.

Optimal pilot design and channel acquisition for conventional MIMO-OFDM has been extensively investigated in the literature. The most common approach is to estimate the channel response in the delay domain, and optimal pilots sent from different transmit antennas are typically assumed to satisfy the phase shift orthogonality condition in both the single-user case \cite{Li02Simplified,Barhumi03Optimal,Minn06Optimal} and the multi-user case \cite{Chi11Training}. Note that such phase shift orthogonal pilots (PSOPs) have been adopted in LTE \cite{Dahlman11LTE}. When channel spatial correlations are taken into account, optimal pilot design has been investigated for both the single-user case \cite{Tuan10Optimized} and multi-user case \cite{Tran12Training}. Although these orthogonal pilot approaches can eliminate pilot interference in the same cell, they do not take into account the pilot overhead issue, which is thought to be one of the limiting factors for throughput in massive MIMO-OFDM [4]. When such approaches are directly adopted in time-division duplex (TDD) massive MIMO-OFDM, the corresponding pilot overhead is proportional to the sum of the number of UT antennas, and would be prohibitively large as the number of UTs becomes large. This becomes the system bottleneck, especially in high mobility scenarios where pilots must be transmitted more frequently. Therefore, a pilot approach that takes the pilot overhead issue into account is of importance for massive MIMO-OFDM systems.

In this paper, we propose adjustable phase shift pilots (APSPs) for massive MIMO-OFDM to reduce the pilot overhead. For APSPs, one sequence along with different adjustable phase shifted versions of itself in the frequency domain are adopted as pilots for different UTs. The proposed APSPs are different from conventional PSOPs \cite{Li02Simplified,Barhumi03Optimal,Chi11Training}, in which phase shifts for different pilots are fixed, and phase shift differences between different pilots are no less than the maximum channel delay (divided by the system sampling duration) of all the UTs. Since in our approach the phase shifts for different pilots are adjustable, more pilots are available compared with conventional PSOPs, which leads to significantly reduced pilot overhead.

The proposed APSPs exploit the following two channel properties: First, wireless channels are sparse in many typical propagation scenarios; most channel power is concentrated in a finite region of delays and/or angles due to limited scattering \cite{Tse05Fundamentals,Clerckx13MIMO,win2chanmod,Bajwa10Compressed}. Such channel sparsity can be resolved in the angle domain in massive MIMO due to the relatively large antenna array apertures, which has been observed in recent massive MIMO channel measurement results \cite{Payami12Channel,Gao13Massive}. Second, channel sparsity patterns, i.e., channel power distributions in the angle-delay domain, for different UTs are usually different.\footnote{There has been recent work that considers channels with a sparse common support \cite{Barbotin12Estimation,Rao14Distributed}. However, for massive MIMO channels, the common support assumption might not hold due to the increased angle resolution \cite{Masood15Efficient,Barbotin12Estimation}. Thus, in this work we assume that the channel sparsity patterns of different UTs are different (but not necessarily totally different), although the proposed APSP approach can also be applied to the common support cases.} For APSPs, when the phase shifts for pilots employed by different UTs are properly scheduled according to the above channel properties, channel acquisition can be achieved simultaneously in an almost interference-free manner as with conventional PSOPs. There has recently been increased research interest on utilizing channel sparsity for channel acquisition in massive MIMO. For instance, a time-frequency training scheme \cite{Dai13Spectrally} and a distributed Bayesian channel estimation scheme \cite{Masood15Efficient} were proposed for massive MIMO-OFDM by exploiting the channel sparsity. As the approaches in \cite{Masood15Efficient} and \cite{Dai13Spectrally} focus on channel acquisition for a single UT, the corresponding pilot overhead would still grow linearly with the number of UTs. Channel sparsity has also been exploited to mitigate pilot contamination in multi-cell massive MIMO \cite{Chen14Pilot,Wen15Channel}.  Note that compressive sensing has been applied to sparse channel acquisition in some recent works (see, e.g., \cite{Bajwa10Compressed,Barbotin12Estimation,Berger10Application,Rao14Distributed} and references therein), in which the corresponding pilot signals are usually assumed to be randomly generated. However, it is usually quite difficult to implement random pilot signals in practical systems \cite{Candes07Sparsity}. For example, adopting large dimensional random pilot signals in the massive MIMO-OFDM systems considered here requires huge storage space and high complexity channel acquisition algorithms. In addition, a low peak-to-average power ratio (PAPR) for randomly generated pilot signals usually cannot be guaranteed. These drawbacks can be mitigated via proper design of the deterministic sensing matrices (see, e.g., \cite{Calderbank10Construction,Strohmer12Measure} and references therein).

The main contributions of this paper are summarized as follows:
\begin{itemize}
\item Based on a physically motivated channel model, we establish a relationship between the space-frequency domain channel covariance matrix (SFCCM) and the channel power angle-delay spectrum for massive MIMO-OFDM. We show that when the number of BS antennas is sufficiently large, the eigenvectors of the SFCCMs for different UTs tend to be equal, while the eigenvalues depend on the respective channel power angle-delay spectra, which reveals the channel sparsity in the angle-delay domain. Then we propose the angle-delay domain channel response matrix (ADCRM) and the corresponding angle-delay domain channel power matrix (ADCPM), which can model the massive MIMO-OFDM channel sparsity in the angle-delay domain, and are convenient for further analyses.
\item With the presented channel model, we propose APSP-based channel acquisition (APSP-CA) for massive MIMO-OFDM in TDD mode. For APSPs, equivalent channels for different UTs will experience corresponding cyclic shifts in the delay domain. Using this property, we show that the sum mean square error (MSE) of channel estimation (MSE-CE) can be minimized if the UTs' channel power distributions in the angle-delay domain can be made non-overlapping with proper pilot phase shift scheduling. Taking the time-varying nature of the channel into account, we further investigate channel prediction during the data segment using the received pilot signals. We show that the sum MSE of channel prediction (MSE-CP) can also be minimized if the UTs' channel power distributions in the angle-delay domain can be made non-overlapping with proper pilot phase shift scheduling, which coincides with the optimal channel estimation condition. A simplified pilot phase shift scheduling algorithm is developed based on this optimal channel acquisition condition. The proposed APSP-CA approach is investigated for cases involving both one symbol and multiple consecutive symbols.
\item The proposed APSP-CA is evaluated in several typical propagation scenarios, and significant performance gains in terms of achievable spectral efficiency over the conventional PSOP-based channel acquisition (PSOP-CA) are demonstrated, especially in high mobility scenarios.
\end{itemize}
Portions of this work previously appeared in the conference paper \cite{You15Adjustable}.

\subsection{Notations}\label{sec:prwb_not}
We adopt the following notation throughout the paper. We use $\barjmath=\sqrt{-1}$ to denote the imaginary unit. $\left\lfloor x\right\rfloor$ ($\left\lceil x\right\rceil$) denotes the largest (smallest) integer not greater (smaller) than $x$. $\modulo{\cdot}{N}$ denotes the modulo-$N$ operation. $\delta(\cdot)$ denotes the delta function. Upper (lower) case boldface letters denote matrices (column vectors). The notation $\triangleq$ is used for definitions. Notations $\sim$ and $\propto$ represent ``distributed as'' and ``proportional to'', respectively. We adopt ${\bI}_{N}$ to denote the $N\times N$ dimensional identity matrix, and $\bI_{N\times G}$ to denote the matrix composed of the first $G\left(\leq N\right)$ columns of $\bI_{N}$. We adopt $\bzero$ to denote the all-zero vector or matrix. The superscripts $(\cdot)^{H}$, $(\cdot)^{T}$, and $(\cdot)^{*}$ denote the conjugate-transpose, transpose, and conjugate operations, respectively. The operator $\diag{\bx}$ denotes the diagonal matrix with $\bx$ along its main diagonal. We employ $\vecele{\ba}{i}$, $\vecele{\bA}{i,j}$ and $\vecele{\bA}{i,:}$ to denote the \ith{i} element of the vector $\ba$, the $(i,j)$th element of the matrix $\bA$ and the \ith{i} row of the matrix $\bA$, respectively, where the element indices start with 0. $\bbC^{M\times N}$ ($\bbR^{M\times N}$) denotes the $M\times N$ dimensional complex (real) vector space. $\expect{\cdot}$ denotes the expectation operation. $\GCN{\ba}{\bB}$ denotes the circular symmetric complex Gaussian distribution with mean $\ba$ and covariance $\bB$. $\otimes$ and $\odot$ denote the Kronecker product and Hadamard product, respectively. $\vecop{\cdot}$ represents the vectorization operation. $\bF_{N}$ denotes the $N$-dimensional unitary discrete Fourier transform (DFT) matrix. $\bF_{N\times G}$ denotes the matrix composed of the first $G\left(\leq N\right)$ columns of $\bF_{N}$. $\bbf_{N,q}$ denotes the \ith{q} column of the matrix $\sqrt{N}\bF_{N}$. We further define the permutation matrix $\bPi_{N}^{n} \triangleq \left[\begin{IEEEeqnarraybox*}[][c]{,c/c,}
\bzero&\bI_{N-\modulo{n}{N}}\\
\bI_{\modulo{n}{N}}&\bzero%
\end{IEEEeqnarraybox*}\right]$. The notation $\backslash$ denotes the set subtraction operation.

\subsection{Outline}
The rest of the paper is organized as follows. In \secref{sec:massive_channel}, we investigate the sparse nature of the massive MIMO-OFDM channel model. In \secref{sec:one_tr}, we propose APSP-CA over one OFDM symbol in massive MIMO-OFDM, including channel estimation and prediction. We investigate the multiple consecutive pilot symbol case in \secref{sec:mul_tr}. Simulation results are presented in \secref{sec:sim_res}, and conclusions are given in \secref{sec:conc_pw}.

\section{Massive MIMO-OFDM Channel Model}\label{sec:massive_channel}

In this section, we propose a physically motivated massive MIMO-OFDM channel model, and investigate the inherent channel sparsity property. We consider a single-cell TDD wideband massive MIMO wireless system which consists of one BS equipped with $M$ antennas and $K$ single-antenna UTs. We denote the UT set as $\cK=\left\{\zeroldots{K-1}\right\}$ where $k\in\cK$ represents the UT index. We assume that the channels of different UTs are statistically independent. We assume that the BS is equipped with a one-dimensional uniform linear array (ULA),\footnote{We adopt the ULA model in this paper for clarity, although our work can be readily extended to more general antenna array models using the techniques in \cite{You15Pilot}.} with antennas separated by one-half wavelength. Then the BS array response vector corresponding to the incidence angle $\theta$ with respect to the perpendicular to the array is given by \cite{Clerckx13MIMO}
\begin{align}\label{eq:ula steer vec}
  \vtheta&=\bigg[1\quad\expx{-\barjmath\pi\sintheta}\quad\ldots\ntb
  &\qquad\ldots\quad\expx{-\barjmath\pi(M-1)\sintheta}\bigg]^{T}\in\bbC^{M\times 1}.
\end{align}
We assume that the signals seen at the BS are constrained to lie in the angle interval $\cA=[-\pi/2,\pi/2]$, which can be achieved through the use of directional antennas at the BS, and thus no signal is received at the BS for incidence angles $\theta\notin\cA$ \cite{You15Pilot}.

We consider OFDM modulation with $\dnnot{N}{c}$ subcarriers, performed via the $\dnnot{N}{c}$-point inverse DFT operation, appended with a guard interval (a.k.a. cyclic prefix) of length $\dnnot{N}{g}\left(\leq\dnnot{N}{c}\right)$ samples. We employ $\dnnot{T}{sym}=\left(\dnnot{N}{c}+\dnnot{N}{g}\right)\dnnot{T}{s}$ and $\dnnot{T}{c}=\dnnot{N}{c}\dnnot{T}{s}$ to denote the OFDM symbol duration with and without the guard interval, respectively, where $\dnnot{T}{s}$ is the system sampling duration \cite{Dahlman11LTE}. We assume that the guard interval length $\dnnot{T}{g}=\dnnot{N}{g}\dnnot{T}{s}$ is longer than the maximum channel delay of all the UTs \cite{Edfors98OFDM,Li98Robust}.

We assume that the channels remain constant during one OFDM symbol, and evolve from symbol to symbol. We denote the uplink (UL) channel gain between the antenna of the \ith{k} UT and the \ith{m} antenna of the BS over OFDM symbol $\ell$ and subcarrier $n$ as $\vecele{\bg_{k,\ell,n}}{m}$. Using a physical channel modeling approach (see, e.g., \cite{Clerckx13MIMO,Liu03Capacity,Barriac04Characterizing,Auer12MIMO,Fleury00First}), the channel response vector $\bg_{k,\ell,n}\in\bbC^{M\times1}$ can be described as
\begin{align}\label{eq:wb_cha_mod}
  \bg_{k,\ell,n}
  &=\sum_{q=0}^{\dnnot{N}{g}-1}\int\limits_{-\infty}^{\infty}\int\limits_{-\pi/2}^{\pi/2}\!\vtheta\cdot\expx{-\barjmath2\pi \frac{n}{\dnnot{T}{c}}\tau}\ntb
  &\quad\cdot\expx{\barjmath2\pi \nu\ell\dnnot{T}{sym}}\cdot
  g_{k}\left(\theta,\tau,\nu\right)\cdot\delfunc{\tau-q\dnnot{T}{s}} \intdx{\theta}\intdx{\nu}\ntb
  &=\sum_{q=0}^{\dnnot{N}{g}-1}\int\limits_{-\infty}^{\infty}\int\limits_{-\pi/2}^{\pi/2}\!\vtheta\cdot\expx{-\barjmath2\pi \frac{n}{\dnnot{N}{c}}q}\ntb
  &\quad\cdot\expx{\barjmath2\pi \nu\ell\dnnot{T}{sym}}\cdot
  g_{k}\left(\theta,q\dnnot{T}{s},\nu\right)\intdx{\theta}\intdx{\nu}
\end{align}
where $\vtheta$ is given in \eqref{eq:ula steer vec}, $g_{k}\left(\theta,\tau,\nu\right)$ is the complex-valued joint angle-delay-Doppler channel gain function of UT $k$ corresponding to the incidence angle $\theta$, delay $\tau$, and Doppler frequency $\nu$. Note that the number of significant channel taps in the delay domain is usually limited, and smaller than $\dnnot{N}{g}$; i.e., $\abs{g_{k}\left(\theta,q\dnnot{T}{s},\nu\right)}$ is approximately $0$ for most $q$. Since the locations of the significant channel taps in the delay domain are usually different for different UTs, we adopt \eqref{eq:wb_cha_mod} in this paper to obtain a general channel representation applicable for all the UTs.

We write the \ith{k} UT's channel at OFDM symbol $\ell$ over all subcarriers as
\begin{equation}\label{eq:chgkl}
  \bG_{k,\ell}=
  \left[\bg_{k,\ell,0}\quad\bg_{k,\ell,1}\quad\ldots\quad\bg_{k,\ell,\dnnot{N}{c}-1}\right]
  \in\bbC^{M\times\dnnot{N}{c}}
\end{equation}
which will be referred to as the space-frequency domain channel response matrix (SFCRM).
From \eqref{eq:wb_cha_mod}, it is not hard to show that
\begin{align}\label{eq:phych}
  \vecop{\bG_{k,\ell}}&=\sum_{q=0}^{\dnnot{N}{g}-1}\int\limits_{-\infty}^{\infty}\int\limits_{-\pi/2}^{\pi/2}\!
  \left[\bbf_{\dnnot{N}{c},q}\otimes\vtheta\right]\cdot\expx{\barjmath2\pi \nu\ell\dnnot{T}{sym}}\ntb
  &\quad\cdot g_{k}\left(\theta,q\dnnot{T}{s},\nu\right)\intdx{\theta}\intdx{\nu}.
\end{align}

We assume that channels with different incidence angles, delays, and/or Doppler frequencies are uncorrelated \cite{Clerckx13MIMO,Auer12MIMO,Fleury00First}. We also assume that the temporal correlations and joint space-frequency domain correlations of the channels can be separated \cite{Li98Robust,Auer12MIMO}, i.e.,
\begin{align}\label{eq:uncorcha}
  &\expect{g_{k}\left(\theta,\tau,\nu\right)g_{k}^{*}\left(\theta',\tau',\nu'\right)}\ntb
  &=\upnot{\ttS_{k}}{ADD}\left(\theta,\tau,\nu\right)\cdot\delfunc{\theta-\theta'}\delfunc{\tau-\tau'}\delfunc{\nu-\nu'}\ntb
  &=\upnot{\ttS_{k}}{AD}\left(\theta,\tau\right)\cdot\upnot{\ttS_{k}}{Dop}\left(\nu\right)\cdot\delfunc{\theta-\theta'}\delfunc{\tau-\tau'}\delfunc{\nu-\nu'}
\end{align}
where $\upnot{\ttS_{k}}{ADD}\left(\theta,\tau,\nu\right)$, $\upnot{\ttS_{k}}{AD}\left(\theta,\tau\right)$, and $\upnot{\ttS_{k}}{Dop}\left(\nu\right)$ represent the power angle-delay-Doppler spectrum, power angle-delay spectrum, and power Doppler spectrum of UT $k$, respectively \cite{Patzold12Mobile,Clerckx13MIMO}.

From \eqref{eq:phych} and \eqref{eq:uncorcha}, we can obtain the following channel statistical property (see \appref{app:der_sta} for the derivations)
\begin{align}\label{eq:expbgkldel}
  \expect{\vecop{\bG_{k,\ell+\Delta_{\ell}}}\vecherop{\bG_{k,\ell}}}=\varrho_{k}\left(\Delta_{\ell}\right)\cdot\bR_{k}
\end{align}
where $\varrho_{k}\left(\Delta_{\ell}\right)$ is the channel temporal correlation function (TCF) given by
\begin{equation}
  \varrho_{k}\left(\Delta_{\ell}\right)\triangleq\int\limits_{-\infty}^{\infty}\!\expx{\barjmath 2\pi\nu\Delta_{\ell}\dnnot{T}{sym}}\cdot\upnot{\ttS_{k}}{Dop}\left(\nu\right)\intdx{\nu}
\end{equation}
and $\bR_{k}$ is the space-frequency domain channel covariance matrix (SFCCM) given by
\begin{align}\label{eq:spa_fre_cov}
  \bR_{k}&\triangleq\sum_{q=0}^{\dnnot{N}{g}-1}\int\limits_{-\pi/2}^{\pi/2}\!
  \left[\bbf_{\dnnot{N}{c},q}\otimes\vtheta\right]
  \left[\bbf_{\dnnot{N}{c},q}\otimes\vtheta\right]^{H}\ntb
  &\qquad\cdot\upnot{\ttS_{k}}{AD}\left(\theta,q\dnnot{T}{s}\right)
  \intdx{\theta}\in\bbC^{M\dnnot{N}{c}\times M\dnnot{N}{c}}.
\end{align}

In this work, we consider the widely accepted Clarke-Jakes channel power Doppler spectrum,\footnote{Although the waves impinging on the BS are assumed to be sparsely distributed in the angle domain due to limited scattering around the BS (typically mounted at an elevated position), the waves departing the mobile UTs are usually uniformly distributed in angle of departure. Thus the Clarke-Jakes spectrum is suitable to model the time variation of the channel \cite{Jakes94Microwave,Patzold12Mobile}.} with the corresponding channel TCF given by \cite{Jakes94Microwave,Patzold12Mobile}
\begin{equation}\label{eq:timcor}
  \varrho_{k}\left(\Delta_{\ell}\right)=\mathrm{J}_{0}\left(2\pi\nu_{k}\dnnot{T}{sym}\Delta_{\ell}\right)
\end{equation}
where $\mathrm{J}_{0}\left(\cdot\right)$ is the zeroth-order Bessel function of the first kind, and $\nu_{k}$ is the Doppler frequency of UT $k$. Note that the Clarke-Jakes power Doppler spectrum is an even function, i.e., $\varrho_{k}(\Delta_{\ell})=\varrho_{k}(-\Delta_{\ell})$, and satisfies $\varrho_{k}\left(0\right)=1$. Also, we assume that according to the law of large numbers, the channel elements exhibit a joint Gaussian distribution, i.e., $\vecop{\bG_{k,\ell}}\sim\GCN{\bzero}{\bR_{k}}$.

Before proceeding, we investigate in the following proposition a property of the large dimensional SFCCM, and present a relationship between the SFCCM and the power angle-delay spectrum for massive MIMO-OFDM channels.

\begin{prop}\label{prop:Decomp_cov}
Define $\bV_{M}\in\bbC^{M\times M}$ as $\vecele{\bV_{M}}{i,j}\triangleq\frac{1}{\sqrt{M}}\cdot\expx{-\barjmath2\pi\frac{i\left(j-M/2\right)}{M}}$, and $\bOmega_{k}\in\bbR^{M\times\dnnot{N}{g}}$ as
\begin{align}\label{eq:cov_eigen_ele}
  \vecele{\bOmega_{k}}{i,j}\triangleq M\dnnot{N}{c}\left(\theta_{i+1}-\theta_{i}\right)\cdot \upnot{\ttS_{k}}{AD}\left(\theta_{i},\tau_{j}\right)
\end{align}
where $\theta_{m}\triangleq\arcsin\left(2m/M-1\right)$, and $\tau_{n}\triangleq n\dnnot{T}{s}$. Then when the number of antennas $M\to\infty$, the SFCCM $\bR_{k}$ tends to $\left(\bF_{\dnnot{N}{c}\times\dnnot{N}{g}}\otimes\bV_{M}\right)\diag{\vecop{\bOmega_{k}}}\left(\bF_{\dnnot{N}{c}\times\dnnot{N}{g}}\otimes\bV_{M}\right)^{H}$
in the sense that, for fixed non-negative integers $i$ and $j$,
\begin{align}\label{eq:Decomp_cov}
  \lim_{\substack{\toinf{M}}}&\Big[\bR_{k}
  -\left(\bF_{\dnnot{N}{c}\times\dnnot{N}{g}}\otimes\bV_{M}\right)
  \diag{\vecop{\bOmega_{k}}}\ntb
  &\quad\cdot\left(\bF_{\dnnot{N}{c}\times\dnnot{N}{g}}\otimes\bV_{M}\right)^{H}\Big]_{i,j}=0.
\end{align}
\end{prop}
\begin{IEEEproof}
See \appref{app:prop_Decomp_cov}.
\end{IEEEproof}

The relationship between the space-frequency domain channel joint correlation property and the channel power distribution in the angle-delay domain for massive MIMO-OFDM is established in \propref{prop:Decomp_cov}. Specifically, for massive MIMO-OFDM channels in the asymptotically large array regime, the eigenvectors of the SFCCMs for different UTs tend to be the same, which shows that massive MIMO-OFDM channels can be asymptotically decorrelated by the fixed space-frequency domain statistical eigendirections, while the eigenvalues depend on the corresponding channel power angle-delay spectra.

\propref{prop:Decomp_cov} indicates that, for massive MIMO-OFDM channels, when the number of BS antennas $M$ is sufficiently large, the SFCCM can be well approximated by
\begin{align}\label{eq:cov_appr}
  \bR_{k}&\approx\left(\bF_{\dnnot{N}{c}\times\dnnot{N}{g}}\otimes\bV_{M}\right)\diag{\vecop{\bOmega_{k}}}\ntb
  &\qquad\cdot\left(\bF_{\dnnot{N}{c}\times\dnnot{N}{g}}\otimes\bV_{M}\right)^{H}.
\end{align}
It is worth noting that the approximation in \eqref{eq:cov_appr} is consistent with existing results in the literature. For frequency-selective single-input single-output channels, \eqref{eq:cov_appr} agrees with the results in \cite{Li98Robust,van95channel}. For frequency-flat massive MIMO channels, the approximation given in \eqref{eq:cov_appr} has been shown to be accurate enough for a practical number of antennas, which usually ranges from 64 to 512 \cite{Wen15Channel,Yin13coordinated,Adhikary13Joint,You15Pilot}, and a detailed numerical example can be found in \cite{Wen15Channel}. Since the SFCCM model given in \eqref{eq:cov_appr} is a good approximation to the more complex physical channel model in \eqref{eq:spa_fre_cov} when the number of BS antennas is sufficiently large, we will thus exclusively use the simplified SFCCM model in \eqref{eq:cov_appr} in the rest of the paper.

Realistic wireless channels are usually not wide-sense stationary \cite{Clerckx13MIMO}, i.e., $\bR_{k}$ varies as time evolves, although with a relatively large time scale.\footnote{The degree of channel stationarity depends on the propagation scenarios. In typical scenarios, the channel statistics vary on the order of seconds \cite{Liu15Two}, while the OFDM symbol length is usually on the order of millisecond \cite{3gpp.36.211}.} In practice, acquisition of the large dimensional $\bR_{k}$ is rather difficult and resource-intensive for massive MIMO-OFDM. However, when we shift our focus from the space-frequency domain to the angle-delay domain, the problem can be significantly simplified. Motivated by the eigenvalue decomposition of the SFCCM given in \eqref{eq:cov_appr}, we decompose the SFCRM as follows
\begin{equation}\label{eq:gkl}
  \bG_{k,\ell}=\bV_{M}\bH_{k,\ell}\bF_{\dnnot{N}{c}\times\dnnot{N}{g}}^{T}
\end{equation}
where
\begin{equation}\label{eq:adcr}
  \bH_{k,\ell}=\bV_{M}^{H}\bG_{k,\ell}\bF_{\dnnot{N}{c}\times\dnnot{N}{g}}^{*}\in\bbC^{M\times\dnnot{N}{g}}
\end{equation}
is referred to as the angle-delay domain channel response matrix (ADCRM) of UT $k$ at OFDM symbol $\ell$. In the following proposition, we derive a statistical property of the ADCRM.

\begin{prop}\label{prop:adcmsts}
For massive MIMO-OFDM channels, when the number of antennas $M\to\infty$, elements of the ADCRM $\bH_{k,\ell}$ satisfy
\begin{align}\label{eq:exbhk}
  &\expect{\vecele{\bH_{k,\ell+\Delta_{\ell}}}{i,j}\vecele{\bH_{k,\ell}}{i',j'}^{*}}\ntb
  &\qquad=\varrho_{k}\left(\Delta_{\ell}\right)\delfunc{i-i'}\delfunc{j-j'}\cdot\vecele{\bOmega_{k}}{i,j}
\end{align}
where $\bOmega_{k}$ is given in \eqref{eq:cov_eigen_ele}.
\end{prop}
\begin{IEEEproof}
See \appref{app:prop_adcmsts}.
\end{IEEEproof}

\propref{prop:adcmsts} shows that, for massive MIMO-OFDM channels, different elements of the ADCRM $\bH_{k,\ell}$ are approximately mutually statistically uncorrelated, which lends the ADCRM in \eqref{eq:adcr} its physical interpretation. Specifically, different elements of the ADCRM correspond to the channel gains for different incidence angles and delays, which can be resolved in massive MIMO-OFDM with a sufficiently large antenna array aperture. Note that $\vecele{\bOmega_{k}}{i,j}$ corresponds to the average power of $\vecele{\bH_{k}}{i,j}$, and can describe the sparsity of the wireless channels in the angle-delay domain. Hereafter we will refer to $\bOmega_{k}$ as the angle-delay domain channel power matrix (ADCPM) of UT $k$. The dimension of the ADCPM $\bOmega_{k}$ is much smaller than that of the SFCCM $\bR_{k}$, and most elements in $\bOmega_{k}$ are approximately zero due to the channel sparsity. In addition, $\bOmega_{k}$ is composed of the variances of independent angle-delay domain channel elements, and thus can be estimated in an element-wise manner. Therefore, in practice there will be enough resources for one to obtain an estimate of $\bOmega_{k}$ with guaranteed accuracy. In the rest of the paper, we will assume that the ADCPMs of all the UTs are known by the BS.

Before we conclude this section, we define the extended ADCRM as follows
\begin{align}\label{eq:extadcrm}
  \barbH_{k,\ell,\singbrac{\dnnot{N}{c}}}
  &\triangleq\bH_{k,\ell}\bI_{\dnnot{N}{c}\times\dnnot{N}{g}}^{T}\ntb
  &=\left[\bH_{k,\ell}
  \quad\bzero_{M\times\left(\dnnot{N}{c}-\dnnot{N}{g}\right)}\right]
  \in\bbC^{M\times\dnnot{N}{c}}.
\end{align}
Similarly, the extended ADCPM, which corresponds to the power distribution of the extended ADCRM $\barbH_{k,\ell,\singbrac{\dnnot{N}{c}}}$, is defined as
\begin{align}\label{eq:extadcpm}
  \barbOmega_{k,\singbrac{\dnnot{N}{c}}}
  &\triangleq\bOmega_{k}\bI_{\dnnot{N}{c}\times\dnnot{N}{g}}^{T}\ntb
  &=\left[\bOmega_{k}
  \quad\bzero_{M\times\left(\dnnot{N}{c}-\dnnot{N}{g}\right)}\right]
  \in\bbR^{M\times\dnnot{N}{c}}.
\end{align}
Such definitions will be employed to simplify the analyses in the following sections.

\section{Channel Acquisition with APSPs over One Symbol}\label{sec:one_tr}

Based on the sparse massive MIMO-OFDM channel model presented in the previous section, we propose APSP-CA for massive MIMO-OFDM, including channel estimation and prediction. In this section, we first investigate the case where the APSPs are sent over one OFDM symbol, while the multiple symbol case will be investigated in the next section.

\subsection{APSPs over One Symbol}
We assume that all the UTs are synchronized. During the UL pilot segment, namely, the \ith{\ell} OFDM symbol of each frame, all the UTs transmit the scheduled pilots simultaneously, and the space-frequency domain signal received at the BS can be represented as
\begin{align}\label{eq:recsigult}
    \bY_{\ell}
    =\sum_{k'=0}^{K-1}\bG_{k',\ell}\bX_{k'}+\bZ_{\ell}\in\bbC^{M\times\dnnot{N}{c}}
\end{align}
where $\vecele{\bY_{\ell}}{i,j}$ denotes the received pilot signal at the \ith{i} antenna over the \ith{j} subcarrier, $\bG_{k,\ell}$ is the SFCRM defined in \eqref{eq:chgkl}, $\bX_{k}=\diag{\bx_{k}}\in\bbC^{\dnnot{N}{c}\times\dnnot{N}{c}}$ denotes the frequency domain pilot signal sent from the \ith{k} UT, $\bZ_{\ell}$ is the additive white Gaussian noise (AWGN) matrix during the UL pilot segment with elements identically and independently distributed (i.i.d.) as $\GCN{0}{\dnnot{\sigma}{ztr}}$, and $\dnnot{\sigma}{ztr}$ is the noise power.

The proposed APSP over one OFDM symbol for a given UT $k$ is given by
\begin{align}\label{eq:pls}
  \bX_{k}\triangleq\sqrt{\dnnot{\sigma}{xtr}}
  \underbrace{\diag{\bbf_{\dnnot{N}{c},\phi_{k}}}}_{\triangleq\bD_{\phi_{k}}}
  \bX
  ,\quad\phi_{k}=\zeroldots{\dnnot{N}{c}-1}
\end{align}
where $\bX=\diag{\bx}\in\bbC^{\dnnot{N}{c}\times\dnnot{N}{c}}$ satisfying $\bX\bX^{H}=\bI_{\dnnot{N}{c}}$ is the basic pilot matrix shared by all UTs in the same cell, and $\dnnot{\sigma}{xtr}$ is the pilot signal transmit power. The APSP signal given in \eqref{eq:pls} can be seen as a phase shifted version of $\sqrt{\dnnot{\sigma}{xtr}}\bX$ with phase shift $\phi_{k}$ in the frequency domain. Note that the proposed APSP has the same PAPR as that of $\bX$ in the time domain, thus existing low PAPR sequence designs can be easily incorporated into our approach. In addition, as the basic pilot matrix $\bX$ can be predetermined, only $\bX$ and the pilot phase shift indices rather than the entire pilot matrices are required to be stored, and the required storage space can be significantly reduced.

From \eqref{eq:pls}, it can be readily obtained that, for $\forall k,k'\in\cK$,
\begin{equation}\label{eq:ijcrocor}
  \bX_{k'}\bX_{k}^{H}
  =\dnnot{\sigma}{xtr}\bD_{\phi_{k'}-\phi_{k}}
\end{equation}
which indicates that cross correlations of the proposed APSPs for different UTs depend only on the associated phase shift difference. It is worth noting that, for conventional PSOPs, the phase shift differences for different pilots are set to satisfy the orthogonality condition $\abs{\phi_{k'}-\phi_{k}}\geq\dnnot{N}{g}\ \forall k'\neq k$. However, for our APSPs, the phase shifts for different pilots are adjustable, and pilots for different UTs can even share the same phase shift, which leads to more available pilots, and thus pilot overhead can be significantly reduced.

\subsection{Channel Estimation with APSPs}

In this section we investigate channel estimation during the pilot segment under the minimum MSE (MMSE) criterion using the proposed APSPs. Direct MMSE estimation of the SFCRM $\bG_{k,\ell}$ requires information about the large dimensional SFCCM $\bR_{k}$ and a large dimensional matrix inversion, which is difficult to implement in practice. However, with the sparse massive MIMO-OFDM channel model presented above, when we shift our focus from the space-frequency domain to the angle-delay domain, channel estimation can be greatly simplified. The BS can first estimate the ADCRM to obtain $\hatbH_{k,\ell}$, then the SFCRM estimates can be readily obtained as $\hatbG_{k,\ell}=\bV_{M}\hatbH_{k,\ell}\bF_{\dnnot{N}{c}\times\dnnot{N}{g}}^{T}$ via exploiting the unitary equivalence between the angle-delay domain channels and the space-frequency domain channels given in \eqref{eq:gkl}, while the same MSE-CE performance can be maintained. In the following, we focus on estimation of the ADCRM $\bH_{k,\ell}$ under the MMSE criterion.

Recalling \eqref{eq:gkl}, the received pilot signal at the BS in \eqref{eq:recsigult} can be rewritten as
\begin{align}\label{eq:recsigad}
    \bY_{\ell}
    =\sum_{k'=0}^{K-1}\bV_{M}\bH_{k',\ell}\bF_{\dnnot{N}{c}\times\dnnot{N}{g}}^{T}
    \bX_{k'}+\bZ_{\ell}.
\end{align}
After decorrelation and power normalization of $\bY_{\ell}$, the BS can obtain an observation of the UL channel $\bH_{k,\ell}$, given by \eqref{eq:ykltr} shown at the top of the next page,
\begin{figure*}
\begin{align}\label{eq:ykltr}
    \bY_{k,\ell}
    &=\oneon{\dnnot{\sigma}{xtr}}\bV_{M}^{H}\bY_{\ell}\bX_{k}^{H}\bF_{\dnnot{N}{c}\times\dnnot{N}{g}}^{*}\ntb
    &=\oneon{\dnnot{\sigma}{xtr}}\sum_{k'=0}^{K-1}\bH_{k',\ell}\bF_{\dnnot{N}{c}\times\dnnot{N}{g}}^{T}\bX_{k'}\bX_{k}^{H}\bF_{\dnnot{N}{c}\times\dnnot{N}{g}}^{*}
    +\oneon{\dnnot{\sigma}{xtr}}\bV_{M}^{H}\bZ_{\ell}\bX_{k}^{H}\bF_{\dnnot{N}{c}\times\dnnot{N}{g}}^{*}\ntb
    &\equaa\bH_{k,\ell}+\underbrace{\sum_{k'\neq k}\bH_{k',\ell}\bF_{\dnnot{N}{c}\times\dnnot{N}{g}}^{T}\bD_{\phi_{k'}-\phi_{k}}\bF_{\dnnot{N}{c}\times\dnnot{N}{g}}^{*}}_{\mathrm{pilot\ interference}\triangleq\sum_{k'\neq k}\bH_{k',\ell}^{\phi_{k'}-\phi_{k}}}
    +\underbrace{\oneon{\dnnot{\sigma}{xtr}}\bV_{M}^{H}\bZ_{\ell}\bX_{k}^{H}\bF_{\dnnot{N}{c}\times\dnnot{N}{g}}^{*}}_{\mathrm{pilot\ noise}}
\end{align}
\hrulefill
\end{figure*}
where (a) follows from \eqref{eq:ijcrocor}. Using the unitary transformation property, it can be readily shown that the pilot noise term in \eqref{eq:ykltr} exhibits a Gaussian distribution with i.i.d. elements distributed as $\GCN{0}{\dnnot{\sigma}{ztr}/\dnnot{\sigma}{xtr}}$, and \eqref{eq:ykltr} can be simplified as
\begin{align}\label{eq:simykltr}
    \bY_{k,\ell}
    =\bH_{k,\ell}+\sum_{k'\neq k}\bH_{k',\ell}^{\phi_{k'}-\phi_{k}}
    +\oneon{\sqrt{\dnnot{\rho}{tr}}}\dnnot{\bZ}{iid}
\end{align}
where $\dnnot{\rho}{tr}\triangleq\dnnot{\sigma}{xtr}/\dnnot{\sigma}{ztr}$ is the signal-to-noise ratio (SNR) during the pilot segment, and $\dnnot{\bZ}{iid}\in\bbC^{M\times\dnnot{N}{g}}$ is the normalized AWGN matrix with  i.i.d. elements distributed as $\GCN{0}{1}$.

Note that the pilot interference term $\bH_{k',\ell}^{\phi_{k'}-\phi_{k}}$ defined in \eqref{eq:ykltr} satisfies
\begin{align}\label{eq:hulphi}
  \bH_{k',\ell}^{\phi_{k'}-\phi_{k}}
  &=\bH_{k',\ell}\bF_{\dnnot{N}{c}\times\dnnot{N}{g}}^{T}\bD_{\phi_{k'}-\phi_{k}}\bF_{\dnnot{N}{c}\times\dnnot{N}{g}}^{*}\ntb
  &=\bH_{k',\ell}\bI_{\dnnot{N}{c}\times\dnnot{N}{g}}^{T}\bF_{\dnnot{N}{c}}^{T}\bD_{\phi_{k'}-\phi_{k}}\bF_{\dnnot{N}{c}}^{*}\bI_{\dnnot{N}{c}\times\dnnot{N}{g}}\ntb
  &\equaa\barbH_{k',\ell,\singbrac{\dnnot{N}{c}}}\bF_{\dnnot{N}{c}}^{T}\bD_{\phi_{k'}-\phi_{k}}\bF_{\dnnot{N}{c}}^{*}\bI_{\dnnot{N}{c}\times\dnnot{N}{g}}\ntb
  &\equab\barbH_{k',\ell,\singbrac{\dnnot{N}{c}}}\bPi_{\dnnot{N}{c}}^{\phi_{k'}-\phi_{k}}\bI_{\dnnot{N}{c}\times\dnnot{N}{g}}
\end{align}
where (a) follows from \eqref{eq:extadcrm}, and (b) follows from the permutation matrix definition given in \secref{sec:prwb_not}. Thus, the pilot interference term $\bH_{k',\ell}^{\phi_{k'}-\phi_{k}}$ in \eqref{eq:simykltr} is a column truncated version of the extended ADCRM $\barbH_{k',\ell,\singbrac{\dnnot{N}{c}}}$ with a cyclic column shift, where the shift factor depends on the corresponding pilot phase shift difference $\phi_{k'}-\phi_{k}$. Thus elements of $\bH_{k',\ell}^{\phi_{k'}-\phi_{k}}$ can be readily obtained as
\begin{align}
  \vecele{\bH_{k',\ell}^{\phi_{k'}-\phi_{k}}}{i,j}
  =\left\{ {\begin{array}{l}
\vecele{\bH_{k',\ell}}{i,\modulo{j-\left(\phi_{k'}-\phi_{k}\right)}{\dnnot{N}{c}}}, \\
 \quad \modulo{j-\left(\phi_{k'}-\phi_{k}\right)}{\dnnot{N}{c}}\leq\dnnot{N}{g}-1 \\
0, \quad\textrm{else}.
\end{array}} \right.
\end{align}

Recalling \propref{prop:adcmsts}, elements of the ADCRM $\bH_{k',\ell}$ are statistically uncorrelated. Consequently, elements of the pilot interference term $\bH_{k',\ell}^{\phi_{k'}-\phi_{k}}$, a column truncated copy of $\bH_{k',\ell}$ with cyclic column shift, are also statistically uncorrelated. Thus, using the same methodology as in the previous section, the corresponding power matrix of the pilot interference term $\bH_{k',\ell}^{\phi_{k'}-\phi_{k}}$ can be defined as
\begin{align}\label{eq:omeuphuk}
  \bOmega_{k'}^{\phi_{k'}-\phi_{k}}
  &\triangleq\expect{\bH_{k',\ell}^{\phi_{k'}-\phi_{k}}\odot\left(\bH_{k',\ell}^{\phi_{k'}-\phi_{k}}\right)^{*}}\ntb
  &=\barbOmega_{k',\singbrac{\dnnot{N}{c}}}\bPi_{\dnnot{N}{c}}^{\phi_{k'}-\phi_{k}}\bI_{\dnnot{N}{c}\times\dnnot{N}{g}}
\end{align}
which is a column truncated version of the extended ADCPM $\barbOmega_{k',\singbrac{\dnnot{N}{c}}}$ defined in \eqref{eq:extadcpm} with cyclic column shift $\phi_{k'}-\phi_{k}$.

With the channel observation $\bY_{k,\ell}$ in \eqref{eq:simykltr}, and the fact that the angle-delay domain channel elements are uncorrelated as derived in \propref{prop:adcmsts}, the MMSE estimate $\hatbH_{k,\ell}$ can be obtained in an element-wise manner as follows \cite{Kailath00Linear}
\begin{align}\label{eq:hathkl}
    \vecele{\hatbH_{k,\ell}}{i,j}=\frac{\vecele{\bOmega_{k}}{i,j}}
  {\sum_{k'=0}^{K-1}\vecele{\bOmega_{k'}^{\phi_{k'}-\phi_{k}}}{i,j}+\oneon{\dnnot{\rho}{tr}}}
  \vecele{\bY_{k,\ell}}{i,j}.
\end{align}
Let $\tildebH_{k,\ell}=\bH_{k,\ell}-\hatbH_{k,\ell}$ be the angle-delay domain channel estimation error of the \ith{k} UT, then the corresponding MSE-CE can be obtained as
\begin{align}\label{eq:epsilonkl}
  \upnot{\epsilon_{k}}{CE}
  &\triangleq\sum_{i=0}^{M-1}\sum_{j=0}^{\dnnot{N}{g}-1}
  \expect{\sqrabs{\vecele{\tildebH_{k,\ell}}{i,j}}}\ntb
  &\equaa\sum_{i=0}^{M-1}\sum_{j=0}^{\dnnot{N}{g}-1}
  \expect{\sqrabs{\vecele{\bH_{k,\ell}}{i,j}}-\sqrabs{\vecele{\hatbH_{k,\ell}}{i,j}}}\ntb
  &=\sum_{i=0}^{M-1}\sum_{j=0}^{\dnnot{N}{g}-1}\left\{\vecele{\bOmega_{k}}{i,j}
  -\frac{\vecele{\bOmega_{k}}{i,j}^{2}}{\sum_{k'=0}^{K-1}\vecele{\bOmega_{k'}^{\phi_{k'}-\phi_{k}}}{i,j}+\oneon{\dnnot{\rho}{tr}}}\right\}
\end{align}
where (a) follows from the orthogonality principle of MMSE estimation \cite{Kailath00Linear}.

Before we proceed, we define the sum MSE-CE of all the UTs as
\begin{align}\label{eq:epsicesum}
  \upnot{{\epsilon}}{CE}\triangleq&\sum_{k=0}^{K-1}\upnot{\epsilon_{k}}{CE}.
\end{align}
Due to the incurred pilot interference, performance of the APSP-based channel estimation might deteriorate. However, we will show in the following proposition that such effects can be eliminated with proper phase shift scheduling for different pilots.

\begin{prop}\label{prop:mmsetrseqcon}
The sum MSE-CE $\upnot{\epsilon}{CE}$ is lower bounded by
\begin{equation}\label{eq:varepsilonl}
  \upnot{\epsilon}{CE}\geq\upnot{\varepsilon}{CE}=\sum_{k=0}^{K-1}\sum_{i=0}^{M-1}\sum_{j=0}^{\dnnot{N}{g}-1}
  \left\{\vecele{\bOmega_{k}}{i,j}-\frac{\vecele{\bOmega_{k}}{i,j}^{2}}{\vecele{\bOmega_{k}}{i,j}+\oneon{\dnnot{\rho}{tr}}}\right\}
\end{equation}
and the lower bound can be achieved under the condition that, for $\forall k,k'\in\cK$ and $k\neq k'$,
\begin{equation}\label{eq:condokou}
  \left(\barbOmega_{k,\singbrac{\dnnot{N}{c}}}\bPi_{\dnnot{N}{c}}^{\phi_{k}}\right)
  \odot\left(\barbOmega_{k',\singbrac{\dnnot{N}{c}}}\bPi_{\dnnot{N}{c}}^{\phi_{k'}}\right)
  =\bzero.
\end{equation}
\end{prop}
\begin{IEEEproof}
See \appref{app:prop_mmsetrseqcon}.
\end{IEEEproof}

\propref{prop:mmsetrseqcon} shows that with the proposed APSPs, the sum MSE-CE can be minimized when phase shifts for different pilots are properly scheduled according to the condition given in \eqref{eq:condokou}. The interpretation is very intuitive. With frequency domain phase shifted pilots, equivalent channels will exhibit corresponding cyclic shifts in the delay domain, as seen from \eqref{eq:hulphi}. If the equivalent channel power distributions in the angle-delay domain for different UTs can be made non-overlapping after pilot phase shift scheduling, the pilot interference effect can be eliminated, and the sum MSE-CE can be minimized.

Wireless channels are approximately sparse in the angle-delay domain in many practical propagation scenarios, and typically only a few elements of the ADCPM $\bOmega_{k}$ are dominant in massive MIMO-OFDM. When such channel sparsity is properly taken into account, the equivalent angle-delay domain channels for different UTs are almost non-overlapping with high probability, assuming proper pilot phase shifts. This suggests the feasibility of the proposed APSPs for massive MIMO-OFDM.

Note that performance of the proposed APSP approach is related to the channel sparsity level. For the case where channels of different UTs have a sparse common support with $s\left(\leq\dnnot{N}{g}\right)$ representing the number of the columns containing non-zero elements in the ADCPM \cite{Barbotin12Estimation,Rao14Distributed}, the maximum number of UTs that can be served without pilot interference is $\left\lfloor\dnnot{N}{c}/s\right\rfloor$. However, for practical wireless channels, most of the channel elements in the angle-delay domain are close to zero, and the condition in \eqref{eq:condokou} usually cannot be satisfied exactly, which will lead to degradation of the channel acquisition performance. In such cases, it is clear that the more sparse the channels are, the better performance can be achieved by the proposed
APSP approach.

Before we conclude this subsection, we remark here that several existing pilot approaches satisfy the optimal condition given in \propref{prop:mmsetrseqcon}. For the case where channel sparsity property is not known, it is reasonable to assume that all the angle-delay domain channel elements are identically distributed, i.e., all the ADCPM elements are equal, in which case the optimal condition in \eqref{eq:condokou} can be achieved when $\abs{\phi_{k}-\phi_{k'}}\geq\dnnot{N}{g}$ for $\forall k\neq k'$, i.e., the extended channels in the delay domain for different UTs are totally separated, which coincides with the conventional PSOPs \cite{Chi11Training}. For frequency-flat massive MIMO channels, i.e., $\dnnot{N}{c}=1$, the condition in \eqref{eq:condokou} can be achieved when $\bOmega_{k}\odot\bOmega_{k'}=\bzero$ for $\forall k\neq k'$, i.e., different UTs can share the same pilot when the respective channels have non-overlapping support in the angle domain, which coincides with previous works such as \cite{You15Pilot,Yin13coordinated}. In our work, the proposed APSPs exploit the joint angle-delay domain channel sparsity in massive MIMO-OFDM, and are more efficient and general from the pilot overhead point of view.

\subsection{Channel Prediction with APSPs}

In the previous subsection, we investigated channel estimation during the pilot segment. Directly employing the pilot segment channel estimates in the data segment might not always be appropriate \cite{Truong13Effects}, especially in high mobility scenarios, which are the main focus of the APSPs. In this subsection, we investigate channel prediction during the data segment based on the received pilot signals, using the proposed APSPs.

For frame-based massive MIMO-OFDM transmission, the BS utilizes the received signals during the pilot segment to acquire the channels in the current frame. If the pilot segment channel estimate $\hatbH_{k,\ell}$ is directly employed as the estimate of the channel $\bH_{k,\ell+\Delta_{\ell}}$ during the data segment, the corresponding sum MSE-CE for a given delay $\Delta_{\ell}$ between the pilot symbol and data symbol can be written as
\begin{align}\label{eq:epscek}
  \upnot{\epsilon}{CE}\left(\Delta_{\ell}\right)
  &=\sum_{k=0}^{K-1}\sum_{i=0}^{M-1}\sum_{j=0}^{\dnnot{N}{g}-1}\expect{\sqrabs{\vecele{\bH_{k,\ell+\Delta_{\ell}}-\hatbH_{k,\ell}}{i,j}}}\ntb
  &=\sum_{k=0}^{K-1}\sum_{i=0}^{M-1}\sum_{j=0}^{\dnnot{N}{g}-1}
  \Bigg\{\vecele{\bOmega_{k}}{i,j}+\left[1-2\varrho_{k}\left(\Delta_{\ell}\right)\right]\ntb
  &\qquad\cdot\frac{\vecele{\bOmega_{k}}{i,j}^{2}}
  {\sum_{k'=0}^{K-1}\vecele{\bOmega_{k'}^{\phi_{k'}-\phi_{k}}}{i,j}+\oneon{\dnnot{\rho}{tr}}}\Bigg\}.
\end{align}
In high mobility scenarios, the channel TCF satisfies $\varrho_{k}\left(\Delta_{\ell}\right)\to0$ for relatively large delay $\abs{\Delta_{\ell}}$. When $\varrho_{k}\left(\Delta_{\ell}\right)<1/2$, i.e., $1-2\varrho_{k}\left(\Delta_{\ell}\right)>0$, it can be observed from \eqref{eq:epscek} that the sum MSE-CE expression $\upnot{\epsilon}{CE}\left(\Delta_{\ell}\right)$ is even larger than the sum channel power $\sum_{k=0}^{K-1}\sum_{i=0}^{M-1}\sum_{j=0}^{\dnnot{N}{g}-1}\vecele{\bOmega_{k}}{i,j}$, and channel estimation performance cannot be guaranteed, which motivates the need for channel prediction.

For channel prediction, the BS utilizes the received pilot signals as well as the channel TCF to get estimates of the channels during the data segment. Under the MMSE criterion, with the angle-delay domain channel property of massive MIMO-OFDM given in \propref{prop:adcmsts}, it is not hard to show that an estimate of the ADCRM $\bH_{k,\ell+\Delta_{\ell}}$ based on $\bY_{k,\ell}$ can be obtained in an element-wise manner as follows
\begin{align}
    &\vecele{\hatbH_{k,\ell+\Delta_{\ell}}}{i,j}\ntb
    &\quad=\varrho_{k}\left(\Delta_{\ell}\right)\frac{\vecele{\bOmega_{k}}{i,j}}
  {\sum_{k'=0}^{K-1}\vecele{\bOmega_{k'}^{\phi_{k'}-\phi_{k}}}{i,j}+\oneon{\dnnot{\rho}{tr}}}\vecele{\bY_{k,\ell}}{i,j}.
\end{align}
Recalling the pilot segment channel estimate in \eqref{eq:hathkl}, it can be seen that
\begin{align}\label{eq:hathkldell}
    \hatbH_{k,\ell+\Delta_{\ell}}=\varrho_{k}\left(\Delta_{\ell}\right)\hatbH_{k,\ell}
\end{align}
which indicates that optimal channel estimates during the data segment can be easily obtained via prediction with initial channel estimates obtained during the pilot segment, and the complexity of channel prediction in massive MIMO-OFDM can be further reduced. Similar to \eqref{eq:epsicesum}, the sum MSE-CP for a given delay $\Delta_{\ell}$ between the data symbol and pilot symbol can be defined as
\begin{align}\label{eq:epsilonklcp}
  \upnot{\epsilon}{CP}\left(\Delta_{\ell}\right)
  &\triangleq\sum_{k=0}^{K-1}\sum_{i=0}^{M-1}\sum_{j=0}^{\dnnot{N}{g}-1}
  \expect{\sqrabs{\vecele{\bH_{k,\ell+\Delta_{\ell}}-\hatbH_{k,\ell+\Delta_{\ell}}}{i,j}}}\ntb
  &=\sum_{k=0}^{K-1}\sum_{i=0}^{M-1}\sum_{j=0}^{\dnnot{N}{g}-1}
  \Bigg\{\vecele{\bOmega_{k}}{i,j}
  -\varrho_{k}^{2}\left(\Delta_{\ell}\right)\ntb
  &\qquad\cdot\frac{\vecele{\bOmega_{k}}{i,j}^{2}}
  {\sum_{k'=0}^{K-1}\vecele{\bOmega_{k'}^{\phi_{k'}-\phi_{k}}}{i,j}
  +\oneon{\dnnot{\rho}{tr}}}\Bigg\}.
\end{align}
From \eqref{eq:epsilonklcp}, it can be seen that pilot interference will affect channel prediction performance similar to the channel estimation case. However, we will show in the following proposition that such effects can still be eliminated with proper pilot phase shift scheduling.
\begin{prop}\label{prop:mmsetrscp}
The sum MSE-CP $\upnot{\epsilon}{CP}\left(\Delta_{\ell}\right)\ \forall\Delta_{\ell}$ is lower bounded by
\begin{align}\label{eq:varepsiloncp}
  &\upnot{\epsilon}{CP}\left(\Delta_{\ell}\right)\geq\upnot{\varepsilon}{CP}\left(\Delta_{\ell}\right)\ntb
  &=\sum_{k=0}^{K-1}\sum_{i=0}^{M-1}\sum_{j=0}^{\dnnot{N}{g}-1}
  \left\{\vecele{\bOmega_{k}}{i,j}-\varrho_{k}^{2}\left(\Delta_{\ell}\right)\frac{\vecele{\bOmega_{k}}{i,j}^{2}}{\vecele{\bOmega_{k}}{i,j}+\oneon{\dnnot{\rho}{tr}}}\right\}
\end{align}
and the lower bound can be achieved under the condition that, for $\forall k,k'\in\cK$ and $k\neq k'$,
\begin{equation}\label{eq:condokoudcp}
  \left(\barbOmega_{k,\singbrac{\dnnot{N}{c}}}\bPi_{\dnnot{N}{c}}^{\phi_{k}}\right)
  \odot\left(\barbOmega_{k',\singbrac{\dnnot{N}{c}}}\bPi_{\dnnot{N}{c}}^{\phi_{k'}}\right)
  =\bzero.
\end{equation}
\end{prop}
\begin{IEEEproof}
The proof is similar to that of \propref{prop:mmsetrseqcon}, and is omitted for brevity.
\end{IEEEproof}

\subsection{Frame Structure}\label{subsec:fra_str}

\begin{figure}[!t]
\centering
\subfloat[Type-A frame structure]{\includegraphics[scale=1]{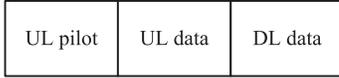}
\label{fig:conframe}}
\vfill
\subfloat[Type-B frame structure]{\includegraphics[scale=1]{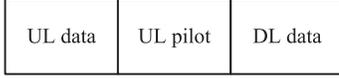}
\label{fig:proframe}}
\caption{Frame structures for TDD transmission.}
\label{fig:frame}
\end{figure}

There exist two typical frame structures for TDD massive MIMO transmission \cite{Bjornson15Massive}. One type of frame structure (which will be referred to as type-A) begins with the UL pilot segment, followed by the UL and downlink (DL) data segments, as shown in \figref{fig:frame}\subref{fig:conframe}. In the second type (which will be referred to as type-B), the UL pilot segment is placed between the UL data segment and DL data segment, as shown in \figref{fig:frame}\subref{fig:proframe}. For the proposed APSP approach, the delay between the tail-end symbols of the data segment and the pilot segment will be longer than the PSOP approach due to the reduced pilot segment length. In addition, the APSP approach focuses on high mobility scenarios where channels vary relatively quickly. Thus the type-B frame structure is well-suited for the proposed APSP approach.

\subsection{Pilot Phase Shift Scheduling}

In the previous subsections, we investigated channel estimation and prediction for massive MIMO-OFDM with APSPs, and obtained the optimal pilot phase shift scheduling condition applicable to both channel estimation and prediction. However, such an optimal condition cannot always be met in practice, but pilot phase shift scheduling can still be beneficial. Several scheduling criteria can be adopted. For example, if we schedule the pilot phase shifts based on the
MMSE-CE criterion, the problem can be formulated as
\begin{align}\label{eq:pspsch}
  \argmin{\left\{\phi_{k}:k\in\cK\right\}}&\qquad\upnot{{\epsilon}}{CE}
\end{align}
where $\upnot{{\epsilon}}{CE}$ is defined in \eqref{eq:epsicesum}. Such a scheduling problem is combinatorial, and optimal solutions must be found through an exhaustive search. Note that the optimal phase shift scheduling conditions for channel estimation and prediction are the same, thus solution of the problem \eqref{eq:pspsch} can also be expected to perform well under the MMSE-CP criterion.

Motivated by the optimal condition for channel estimation and prediction obtained in previous subsections, a simplified pilot phase shift scheduling algorithm can be developed. We first define the following function that measures the degree of overlap between two real matrices $\bA,\bB\in\bbR^{M\times N}$ as follows
\begin{equation}\label{eq:xiab}
  \xi\left(\bA,\bB\right)\triangleq
  \frac{\sum_{i,j}\vecele{\bA\odot\bB}{i,j}}
  {\sqrt{\sum_{i,j}\vecele{\bA}{i,j}^{2}}\cdot
   \sqrt{\sum_{i,j}\vecele{\bB}{i,j}^{2}}}.
\end{equation}
From the Cauchy-Schwarz inequality, it is obvious that the overlapping degree function in \eqref{eq:xiab} satisfies $0\leq\xi\left(\bA,\bB\right)\leq1$. When $\bA$ is a scaled version of $\bB$, $\xi\left(\bA,\bB\right)=1.$ When the locations of non-zero elements in $\bA$ and $\bB$ are non-overlapping, $\xi\left(\bA,\bB\right)=0$. In our algorithm, we preset a threshold to balance the tradeoff between the algorithm complexity and channel acquisition performance. Specifically, we schedule the pilot phase shifts for different UTs to make the overlap function between the ADCPMs for different UTs smaller than the preset threshold $\gamma$. Intuitively, the smaller the threshold $\gamma$, the better the channel acquisition performance will be, although with a higher algorithm complexity. The description of the proposed algorithm is summarized in \alref{alg:GPPSSA}.

\begin{algorithm}[!t]
\caption{Pilot Phase Shift Scheduling Algorithm}
\label{alg:GPPSSA}
\begin{algorithmic}[1]
\Require
The UT set $\cK$ and the corresponding ADCPMs $\left\{\bOmega_{k}:k\in\cK\right\}$; the preset threshold $\gamma$
\Ensure
Pilot phase shift pattern $\left\{\phi_{k}:k\in\cK\right\}$
\State Initialization: $\phi_{0}=0$, scheduled UT set $\upnot{\cK}{sch}=\left\{0\right\}$, unscheduled UT set $\upnot{\cK}{un}=\setdif{\cK}{\upnot{\cK}{sch}}$
\For{$k\in\upnot{\cK}{un}$}
\State Search for a phase shift $\phi$ that satisfies $\xi\left(\barbOmega_{k,\left(\dnnot{N}{c}\right)}\bPi_{\dnnot{N}{c}}^{\phi},
\sum_{k'\in\upnot{\cK}{sch}}\barbOmega_{k',\left(\dnnot{N}{c}\right)}\bPi_{\dnnot{N}{c}}^{\phi_{k'}}\right)\leq\gamma$
\State If $\phi$ cannot be found in step 3, then $\phi=\argmin{x}\ \xi\left(\barbOmega_{k,\left(\dnnot{N}{c}\right)}\bPi_{\dnnot{N}{c}}^{x},
\sum_{k'\in\upnot{\cK}{sch}}\barbOmega_{k',\left(\dnnot{N}{c}\right)}\bPi_{\dnnot{N}{c}}^{\phi_{k'}}\right)$
\State Update $\phi_{k}=\phi$, $\upnot{\cK}{sch}\leftarrow\upnot{\cK}{sch}\cup\left\{k\right\}$, $\upnot{\cK}{un}\leftarrow\setdif{\upnot{\cK}{un}}{\left\{k\right\}}$
\EndFor
\end{algorithmic}
\end{algorithm}

\section{Channel Acquisition with APSPs over Multiple Symbols}\label{sec:mul_tr}

In the previous section, we investigated channel acquisition for massive MIMO-OFDM with the proposed APSPs over one OFDM symbol. Sometimes pilots over one symbol might be not sufficient to accommodate a large number of UTs. In this section, we extend the use of APSPs to the case of multiple consecutive OFDM symbols.

We assume that the pilots are sent over $Q$ consecutive OFDM symbols starting with the \ith{\ell} symbol in each frame. In practice, the pilot segment length $Q$ is usually short, and we adopt the widely accepted assumption that the channels remain constant during the pilot segment \cite{Barhumi03Optimal,Minn06Optimal,Chi11Training}. Then the received signals by the BS during the pilot segment can be written as
\begin{align}\label{eq:ytrlbq}
    \bY_{\ell,\singbrac{Q}}
    &=\sum_{k'=0}^{K-1}\bG_{k',\ell}\bX_{k',\singbrac{Q}}+\bZ_{\ell,\singbrac{Q}}\ntb
    &=\sum_{k'=0}^{K-1}\bV_{M}\bH_{k',\ell}\bF_{\dnnot{N}{c}\times\dnnot{N}{g}}^{T}\bX_{k',\singbrac{Q}}\ntb
    &\qquad+\bZ_{\ell,\singbrac{Q}}\in\bbC^{M\times\dnnot{N}{c}Q}
\end{align}
where $\bY_{\ell,\singbrac{Q}}\triangleq\left[\bY_{\ell}\quad\bY_{\ell+1}\quad\ldots\quad\bY_{\ell+Q-1}\right]$, $\bY_{\ell}\in\bbC^{M\times\dnnot{N}{c}}$ represents the received pilot signal at the BS during the \ith{\ell} symbol, $\bX_{k,\singbrac{Q}}\triangleq\left[\bX_{k,0}\quad\bX_{k,1}\quad\ldots\quad\bX_{k,Q-1}\right]$ represents the pilot signals and $\bX_{k,q}=\diag{\bx_{k,q}}\in\bbC^{\dnnot{N}{c}\times\dnnot{N}{c}}$ represents the signal sent from the \ith{k} UT during the \ith{q} symbol of the pilot segment, $\bZ_{\ell,\singbrac{Q}}$ is AWGN with i.i.d. elements distributed as $\GCN{0}{\dnnot{\sigma}{ztr}}$ and $\dnnot{\sigma}{ztr}$ is the noise power.

Recalling \eqref{eq:pls}, the maximum adjustable phase shift for different pilots over one OFDM symbol is $\dnnot{N}{c}-1$. For the $Q$ pilot symbol case, the maximum adjustable pilot phase shift can be extended to $Q\dnnot{N}{c}-1$. By exploiting the modulo operation, we construct the APSPs over multiple OFDM symbols as follows
\begin{align}\label{eq:xtruq}
  \bX_{k,\singbrac{Q}}
  \triangleq\sqrt{Q}\vecele{\bU}{\modulo{\phi_{k}}{Q},:}
  &\otimes\bX_{\left\lfloor\phi_{k}/Q\right\rfloor},\ntb
  &\qquad\phi_{k}=\zeroldots{Q\dnnot{N}{c}-1}
\end{align}
where $\bU$ is an arbitrary $Q\times Q$ dimensional unitary matrix, and $\bX_{\left\lfloor\phi_{k}/Q\right\rfloor}$ is the APSP signal over one symbol defined in \eqref{eq:pls}. Then it can be obtained that, for $\forall k,k'\in\cK$,
\begin{align}\label{eq:xiqxjq}
&\bX_{k',\singbrac{Q}}\left(\bX_{k,\singbrac{Q}}\right)^{H}\ntb
&\qquad=Q\left(\vecele{\bU}{\modulo{\phi_{k'}}{Q},:}
  \otimes\bX_{\left\lfloor\phi_{k'}/Q\right\rfloor}\right)\ntb
&\qquad\quad\cdot\left(\vecele{\bU}{\modulo{\phi_{k}}{Q},:}
  \otimes\bX_{\left\lfloor\phi_{k}/Q\right\rfloor}\right)^{H}\ntb
&\qquad\equaa Q\left(\vecele{\bU}{\modulo{\phi_{k'}}{Q},:}\vecele{\bU}{\modulo{\phi_{k}}{Q},:}^{H}\right)\otimes
\left(\bX_{\left\lfloor\phi_{k'}/Q\right\rfloor}\bX_{\left\lfloor\phi_{k}/Q\right\rfloor}^{H}\right)\ntb
&\qquad\equab \dnnot{\sigma}{xtr}Q\delfunc{\modulo{\phi_{k'}}{Q}-\modulo{\phi_{k}}{Q}}
\cdot\bD_{\left\lfloor\phi_{k'}/Q\right\rfloor-\left\lfloor\phi_{k}/Q\right\rfloor}
\end{align}
where (a) follows from the Kronecker product identities $\left(\bA\otimes\bB\right)\left(\bC\otimes\bD\right)=\left(\bA\bC\right)\otimes\left(\bB\bD\right)$ and $\left(\bA\otimes\bB\right)^{H}=\bA^{H}\otimes\bB^{H}$ \cite{Seber08Matrix}, and (b) follows from \eqref{eq:ijcrocor}. This shows that the available phase shifts for the $Q$ symbol case are divided into $Q$ groups for the proposed APSPs in \eqref{eq:xtruq}, and the group index depends on the residue of the pilot phase shift $\phi$ with respect to the pilot segment length $Q$. Pilot interference can only affect the UTs using APSPs with phase shifts in the same group. For example, if $\modulo{\phi_{k'}}{Q}=\modulo{\phi_{k}}{Q}$, then phase shifts $\phi_{k'}$ and $\phi_{k}$ are within the same group, and the corresponding channel acquisition of UTs $k'$ and $k$ might be mutually affected.

Given the APSP correlation property over multiple symbols in \eqref{eq:xiqxjq}, the channel estimation and prediction operations can be performed similarly to the single-symbol case investigated in the previous section, and we will briefly discuss such issues below.

After decorrelation and power normalization with $\bY_{\ell,\singbrac{Q}}$ given in \eqref{eq:ytrlbq}, the BS can obtain an observation of the pilot segment ADCRM $\bH_{k,\ell}$ as
\begin{align}\label{eq:yklqtr}
    &\bY_{k,\ell,\singbrac{Q}}\ntb
    &\quad=
    \oneon{\dnnot{\sigma}{xtr}Q}\bV_{M}^{H}\bY_{\ell,\singbrac{Q}}\bX_{k,\singbrac{Q}}^{H}\bF_{\dnnot{N}{c}\times\dnnot{N}{g}}^{*}\ntb
    &\quad=\oneon{\dnnot{\sigma}{xtr}Q}\sum_{k'=0}^{K-1}\bH_{k',\ell}\bF_{\dnnot{N}{c}\times\dnnot{N}{g}}^{T}\bX_{k',\singbrac{Q}}
    \bX_{k,\singbrac{Q}}^{H}\bF_{\dnnot{N}{c}\times\dnnot{N}{g}}^{*}\ntb
    &\qquad+\oneon{\dnnot{\sigma}{xtr}Q}\bV_{M}^{H}\bZ_{\ell,\singbrac{Q}}\bX_{k,\singbrac{Q}}^{H}\bF_{\dnnot{N}{c}\times\dnnot{N}{g}}^{*}\ntb
    &\quad\equaa\sum_{k'=0}^{K-1}\delfunc{\modulo{\phi_{k'}}{Q}-\modulo{\phi_{k}}{Q}}\cdot\bH_{k',\ell}
    \bF_{\dnnot{N}{c}\times\dnnot{N}{g}}^{T}\ntb
    &\qquad\cdot\bD_{\left\lfloor\phi_{k'}/Q\right\rfloor-\left\lfloor\phi_{k}/Q\right\rfloor}
    \bF_{\dnnot{N}{c}\times\dnnot{N}{g}}^{*}
    +\oneon{\sqrt{\dnnot{\rho}{tr}Q}}\dnnot{\bZ}{iid}\ntb
    &\quad\equab\bH_{k,\ell}+\underbrace{\sum_{k'\neq k}\delfunc{\modulo{\phi_{k'}}{Q}-\modulo{\phi_{k}}{Q}}\cdot\bH_{k',\ell}^{\left\lfloor\phi_{k'}/Q\right\rfloor-\left\lfloor\phi_{k}/Q\right\rfloor}}_{\mathrm{pilot\ interference}}\ntb
    &\qquad+\underbrace{\oneon{\sqrt{\dnnot{\rho}{tr}Q}}\dnnot{\bZ}{iid}}_{\mathrm{pilot\ noise}}
\end{align}
where (a) follows from \eqref{eq:xiqxjq}, $\dnnot{\rho}{tr}\triangleq\dnnot{\sigma}{xtr}/\dnnot{\sigma}{ztr}$ is the pilot segment SNR, $\dnnot{\bZ}{iid}$ is the normalized AWGN matrix with i.i.d. elements distributed as $\GCN{0}{1}$, and (b) follows from \eqref{eq:hulphi}.

With the channel observation $\bY_{k,\ell,\singbrac{Q}}$ in \eqref{eq:yklqtr}, the MMSE estimate of the ADCRM $\bH_{k,\ell}$ can be readily obtained in an element-wise manner as \eqref{eq:hathklmul} shown at the top of the next page,
\begin{figure*}
\begin{align}\label{eq:hathklmul}
    \vecele{\hatbH_{k,\ell}}{i,j}=
    \frac{\vecele{\bOmega_{k}}{i,j}}
  {\sum_{k'=0}^{K-1}\delfunc{\modulo{\phi_{k'}}{Q}-\modulo{\phi_{k}}{Q}}
  \vecele{\bOmega_{k'}^{\left\lfloor\phi_{k'}/Q\right\rfloor-\left\lfloor\phi_{k}/Q\right\rfloor}}{i,j}
  +\oneon{\dnnot{\rho}{tr}Q}}\vecele{\bY_{k,\ell,\singbrac{Q}}}{i,j}
\end{align}
\hrule
\end{figure*}
and the corresponding sum MSE-CE is given by \eqref{eq:epsilonklq} shown at the top of the next page,
\begin{figure*}
\begin{align}\label{eq:epsilonklq}
  \upnot{\epsilon_{\singbrac{Q}}}{CE}
  =\sum_{k=0}^{K-1}\sum_{i=0}^{M-1}\sum_{j=0}^{\dnnot{N}{g}-1}\left\{\vecele{\bOmega_{k}}{i,j}
  -\frac{\vecele{\bOmega_{k}}{i,j}^{2}}
   {\sum_{k'=0}^{K-1}\delfunc{\modulo{\phi_{k'}}{Q}-\modulo{\phi_{k}}{Q}}
  \vecele{\bOmega_{k'}^{\left\lfloor\phi_{k'}/Q\right\rfloor-\left\lfloor\phi_{k}/Q\right\rfloor}}{i,j}
  +\oneon{\dnnot{\rho}{tr}Q}}\right\}
\end{align}
\hrule
\end{figure*}
In addition, prediction of the ADCRM $\bH_{k,\ell+\Delta_{\ell}}$ based on $\bY_{k,\ell,\singbrac{Q}}$ can be performed as \eqref{eq:hathkldellmul} shown at the top of the next page,
\begin{figure*}
\begin{align}\label{eq:hathkldellmul}
    \vecele{\hatbH_{k,\ell+\Delta_{\ell}}}{i,j}=
    \frac{\varrho_{k}\left(\Delta_{\ell}\right)\vecele{\bOmega_{k}}{i,j}}
  {\sum_{k'=0}^{K-1}\delfunc{\modulo{\phi_{k'}}{Q}-\modulo{\phi_{k}}{Q}}
  \vecele{\bOmega_{k'}^{\left\lfloor\phi_{k'}/Q\right\rfloor-\left\lfloor\phi_{k}/Q\right\rfloor}}{i,j}
  +\oneon{\dnnot{\rho}{tr}Q}}\vecele{\bY_{k,\ell,\singbrac{Q}}}{i,j}
\end{align}
\hrule
\end{figure*}
and the corresponding sum MSE-CP with a given delay $\Delta_{\ell}$ is given by \eqref{eq:sumcpmul} shown at the top of the next page.
\begin{figure*}
\begin{align}\label{eq:sumcpmul}
  \upnot{\epsilon_{\singbrac{Q}}}{CP}\left(\Delta_{\ell}\right)
  =\sum_{k=0}^{K-1}\sum_{i=0}^{M-1}\sum_{j=0}^{\dnnot{N}{g}-1}\left\{\vecele{\bOmega_{k}}{i,j}
  -\frac{\varrho_{k}^{2}\left(\Delta_{\ell}\right)\vecele{\bOmega_{k}}{i,j}^{2}}
   {\sum_{k'=0}^{K-1}\delfunc{\modulo{\phi_{k'}}{Q}-\modulo{\phi_{k}}{Q}}
  \vecele{\bOmega_{k'}^{\left\lfloor\phi_{k'}/Q\right\rfloor-\left\lfloor\phi_{k}/Q\right\rfloor}}{i,j}
  +\oneon{\dnnot{\rho}{tr}Q}}\right\}
\end{align}
\hrule
\end{figure*}

Based on the above sum MSE-CE and MSE-CP expressions for the multiple symbol APSP case, we can readily obtain the following proposition.
\begin{prop}\label{prop:mmsetrscpmul}
The sum MSE-CE $\upnot{\epsilon_{\singbrac{Q}}}{CE}$ is lower bounded by
\begin{equation}\label{eq:varepsiloncpmul}
  \upnot{\epsilon_{\singbrac{Q}}}{CE}\geq
  \upnot{\varepsilon_{\singbrac{Q}}}{CE}=\sum_{k=0}^{K-1}\sum_{i=0}^{M-1}\sum_{j=0}^{\dnnot{N}{g}-1}
  \left\{\vecele{\bOmega_{k}}{i,j}-\frac{\vecele{\bOmega_{k}}{i,j}^{2}}{\vecele{\bOmega_{k}}{i,j}+\oneon{\dnnot{\rho}{tr}Q}}\right\}
\end{equation}
and the sum MSE-CP $\upnot{\epsilon_{\singbrac{Q}}}{CP}\left(\Delta_{\ell}\right)$ for $\forall\Delta_{\ell}$ is lower bounded by
\begin{align}\label{eq:cpminmul}
  &\upnot{\epsilon_{\singbrac{Q}}}{CP}\left(\Delta_{\ell}\right)\geq
  \upnot{\varepsilon_{\singbrac{Q}}}{CP}\left(\Delta_{\ell}\right)\ntb
  &\qquad=
  \sum_{k=0}^{K-1}\sum_{i=0}^{M-1}\sum_{j=0}^{\dnnot{N}{g}-1}
  \left\{\vecele{\bOmega_{k}}{i,j}-\frac{\varrho_{k}^{2}\left(\Delta_{\ell}\right)\vecele{\bOmega_{k}}{i,j}^{2}}{\vecele{\bOmega_{k}}{i,j}+\oneon{\dnnot{\rho}{tr}Q}}\right\}.
\end{align}
Both the lower bounds in \eqref{eq:varepsiloncpmul} and \eqref{eq:cpminmul} can be achieved under the condition that, for $\forall k,k'\in\cK$ and $k\neq k'$,
\begin{align}\label{eq:condokoudcpmul}
  \left(\barbOmega_{k,\singbrac{\dnnot{N}{c}}}\bPi_{\dnnot{N}{c}}^{\left\lfloor\phi_{k}/Q\right\rfloor}\right)
  &\odot\left(\barbOmega_{k',\singbrac{\dnnot{N}{c}}}\bPi_{\dnnot{N}{c}}^{\left\lfloor\phi_{k'}/Q\right\rfloor}\right)=\bzero,\ntb
  &\qquad\qquad\textrm{when} \quad \modulo{\phi_{k}}{Q}=\modulo{\phi_{k'}}{Q}.
\end{align}
\end{prop}
\begin{IEEEproof}
The proof is similar to that of \propref{prop:mmsetrseqcon}, and is omitted for brevity.
\end{IEEEproof}

\propref{prop:mmsetrscpmul} extends the single-symbol APSP case in the previous section to the multiple symbol case. Actually, when $Q=1$, \propref{prop:mmsetrscpmul} reduces to the results in \propref{prop:mmsetrseqcon} and \propref{prop:mmsetrscp}. The interpretation of \propref{prop:mmsetrscpmul} is straightforward. For multiple symbol APSPs, different pilot phase shifts are divided into several groups, and pilot interference only affects the UTs using the phase shifts within the same group. If pilot interference can be eliminated through proper phase shift scheduling in all the groups, then optimal channel estimation and prediction performance can be achieved. When the optimal pilot phase shift scheduling condition in \propref{prop:mmsetrscpmul} cannot be met, a straightforward extension of the pilot phase shift scheduling algorithm in the previous section can be applied. Specifically, the UT set can be divided into $Q$ groups, and pilot phase shift scheduling can be performed within each UT group using \alref{alg:GPPSSA}. The tradeoff between channel acquisition performance and algorithm complexity can still be balanced with the preset threshold to determine the degree of allowable channel overlap.

\section{Numerical Results}\label{sec:sim_res}

In this section, we present numerical simulations to evaluate the performance of the proposed APSP-CA in massive MIMO-OFDM. The major OFDM parameters, which are based on 3GPP LTE \cite{3gpp.36.211}, are summarized in \tabref{tb:ofdm_para}. The massive MIMO-OFDM system considered is assumed to be equipped with a 128-antenna ULA at the BS with half wavelength antenna spacing. The number of UTs is set to $K=42$ as in \cite{Marzetta10Noncooperative}.

We consider channels with 20 taps in the delay domain, which exhibit an exponential power delay profile \cite{Pedersen00stochastic,win2chanmod}
\begin{align}\label{eq:exp_pdp}
  \upnot{\ttS_{k}}{del}\left(\tau\right)
  \propto\expx{-\tau/\varsigma_{k}},\ \fortext \ \tau\in \left[0,\dnnot{N}{g}\dnnot{T}{s}\right]
\end{align}
where $\varsigma_{k}$ denotes the channel delay spread of UT $k$. We assume that transmissions from all the UTs are synchronized \cite{Dahlman11LTE,win2chanmod}. The \ith{q} channel tap of UT $k$ is assumed to exhibit a Laplacian power angle spectrum \cite{Pedersen00stochastic,win2chanmod,You15Pilot}
\begin{align}\label{eq:lap_pas}
  \upnot{\ttS_{k,q}}{ang}\left(\theta\right)&\propto \expx{-\sqrt{2}\abs{\theta-\theta_{k,q}}/\varphi_{k,q}},\ntb
  &\qquad\qquad \fortext \ \theta\in\cA=[-\pi/2,\pi/2]
\end{align}
where $\theta_{k,q}$ and $\varphi_{k,q}$ represent the corresponding mean angle of arrival (AoA) and angle spread for the given channel tap, respectively. We assume that the UTs are uniformly distributed in a $120^{\circ}$ sector, and the mean AoA $\theta_{k,q}$ is uniformly distributed in the angle interval $[-\pi/3,\pi/3]$ in radians. We do not consider large scale fading in the simulations, and channels are normalized as $\sum_{i,j}\vecele{\bOmega_{k}}{i,j}=M\dnnot{N}{c}$ for $\forall k$. We consider channel propagation under several typical mobility scenarios including suburban (SU), urban macro (UMa), and urban micro (UMi). The primary statistical channel parameters under these scenarios are based on the WINNER II channel model \cite{win2chanmod,Auer12MIMO}, and are summarized in \tabref{tb:cha_sta_par}. We assume that all UTs exhibit the same Doppler, delay, and angle spread in the simulations.

\newcolumntype{L}{>{\hspace*{-\tabcolsep}}l}
\newcolumntype{R}{c<{\hspace*{-\tabcolsep}}}
\definecolor{lightblue}{rgb}{0.93,0.95,1.0}
\begin{table}[!t]
\caption{OFDM System Parameters}\label{tb:ofdm_para}
\centering
\ra{1.3}
\begin{tabular}{LcR}
\toprule
Parameter & & Value\\
\midrule
\rowcolor{lightblue}
System bandwidth && 20 MHz\\
Sampling duration $\dnnot{T}{s}$ && 32.6 ns\\
\rowcolor{lightblue}
Subcarrier spacing && 15 kHz\\
Subcarrier number $\dnnot{N}{c}$ && 2048\\
\rowcolor{lightblue}
Guard interval $\dnnot{N}{g}$ && 144\\
Symbol length $\dnnot{T}{sym}$ && 71.4 $\mu$s\\
\bottomrule
\end{tabular}
\end{table}

\newcolumntype{L}{>{\hspace*{-\tabcolsep}}l}
\newcolumntype{R}{c<{\hspace*{-\tabcolsep}}}
\definecolor{lightblue}{rgb}{0.93,0.95,1.0}
\begin{table}[!t]
\caption{Statistical Channel Parameters in Typical Scenarios}\label{tb:cha_sta_par}
\centering
\ra{1.3}
\begin{tabular}{LcccR}
\toprule
Scenario && \tabincell{c}{Doppler $\nu\dnnot{T}{sym}$ \\ (Velocity)} & \tabincell{c}{Delay \\ spread $\varsigma$} & \tabincell{c}{Angle\\ spread $\varphi$}\\
\midrule
\rowcolor{lightblue}
Suburban && $31\times10^{-3}$ & 0.77 $\mu$s & $2^{\circ}$\\
\rowcolor{lightblue}
(SU) && (\kmh{250}) & & \\
Urban macro  && $14\times10^{-3}$ & 1.85 $\mu$s & $2^{\circ}$\\
(UMa) && (\kmh{100}) &&\\
\rowcolor{lightblue}
Urban micro && $6.6\times10^{-3}$ & 0.62 $\mu$s & $10^{\circ}$\\
\rowcolor{lightblue}
(UMi) && (\kmh{50}) & &\\
\bottomrule
\end{tabular}
\end{table}

\begin{figure*}[!t]
\centering
\subfloat[SU]{\includegraphics[width=0.32\textwidth]{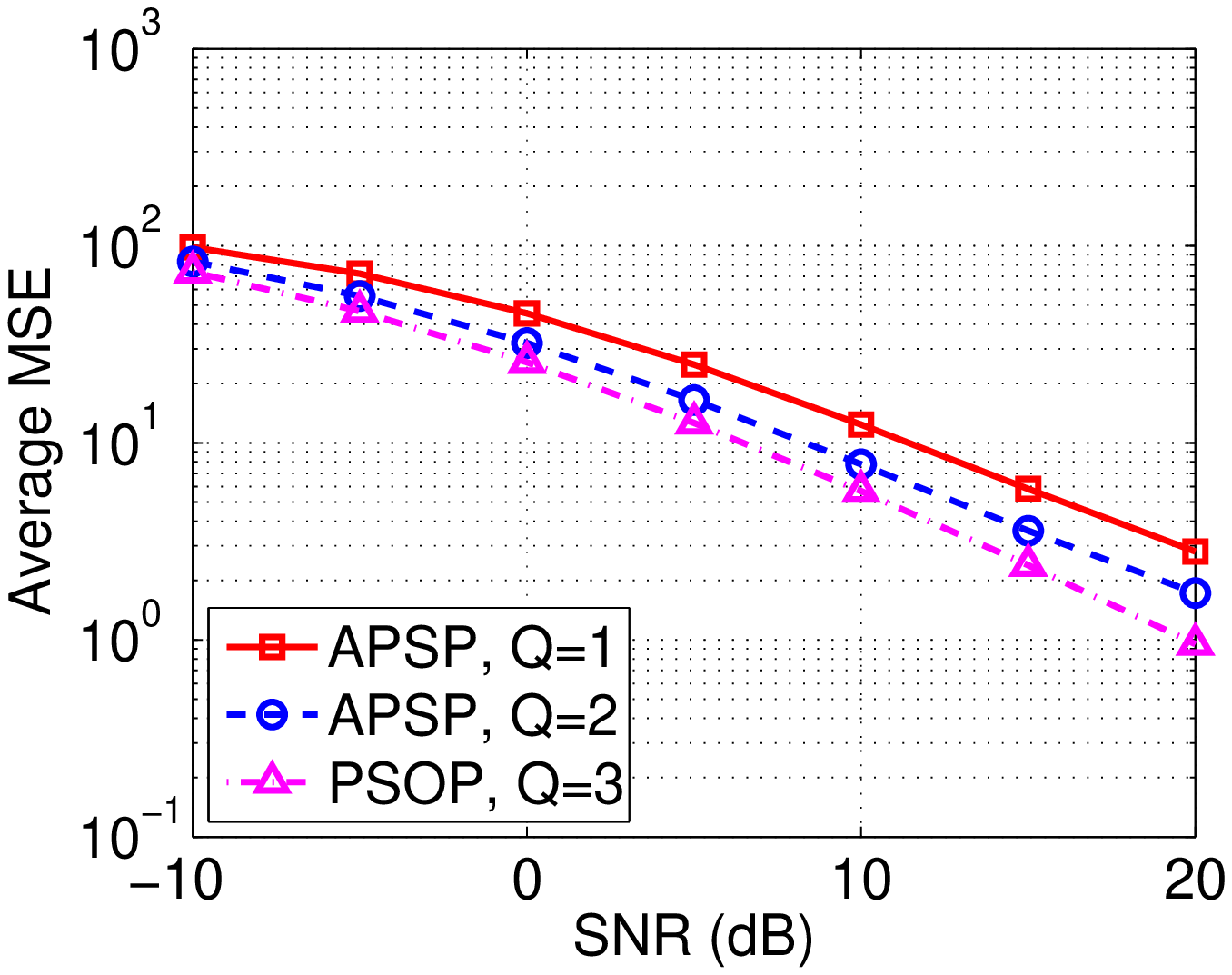}
\label{fig:psmsecoma}}
\hfill
\subfloat[UMa]{\includegraphics[width=0.32\textwidth]{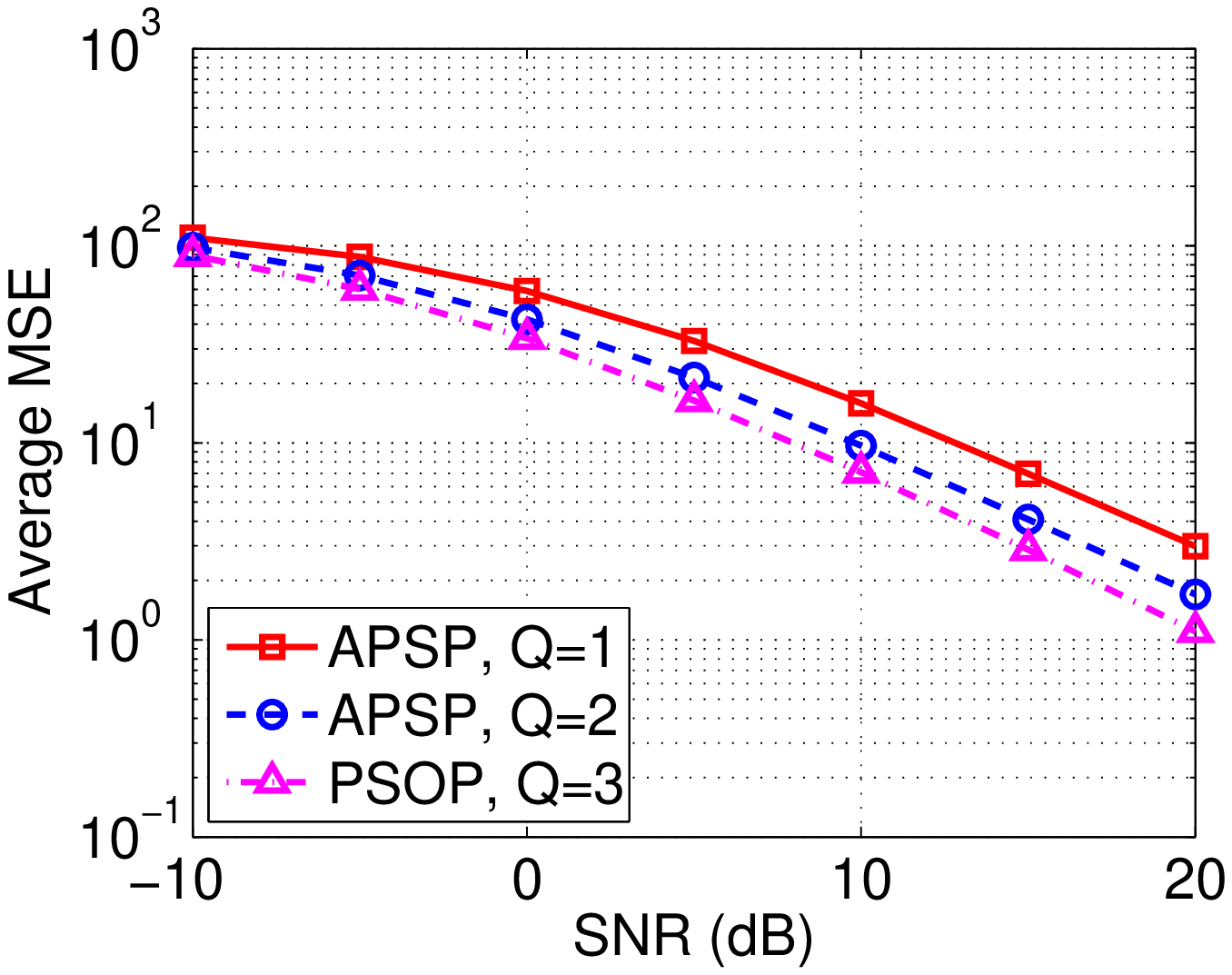}
\label{fig:psmsecomb}}
\hfill
\subfloat[UMi]{\includegraphics[width=0.32\textwidth]{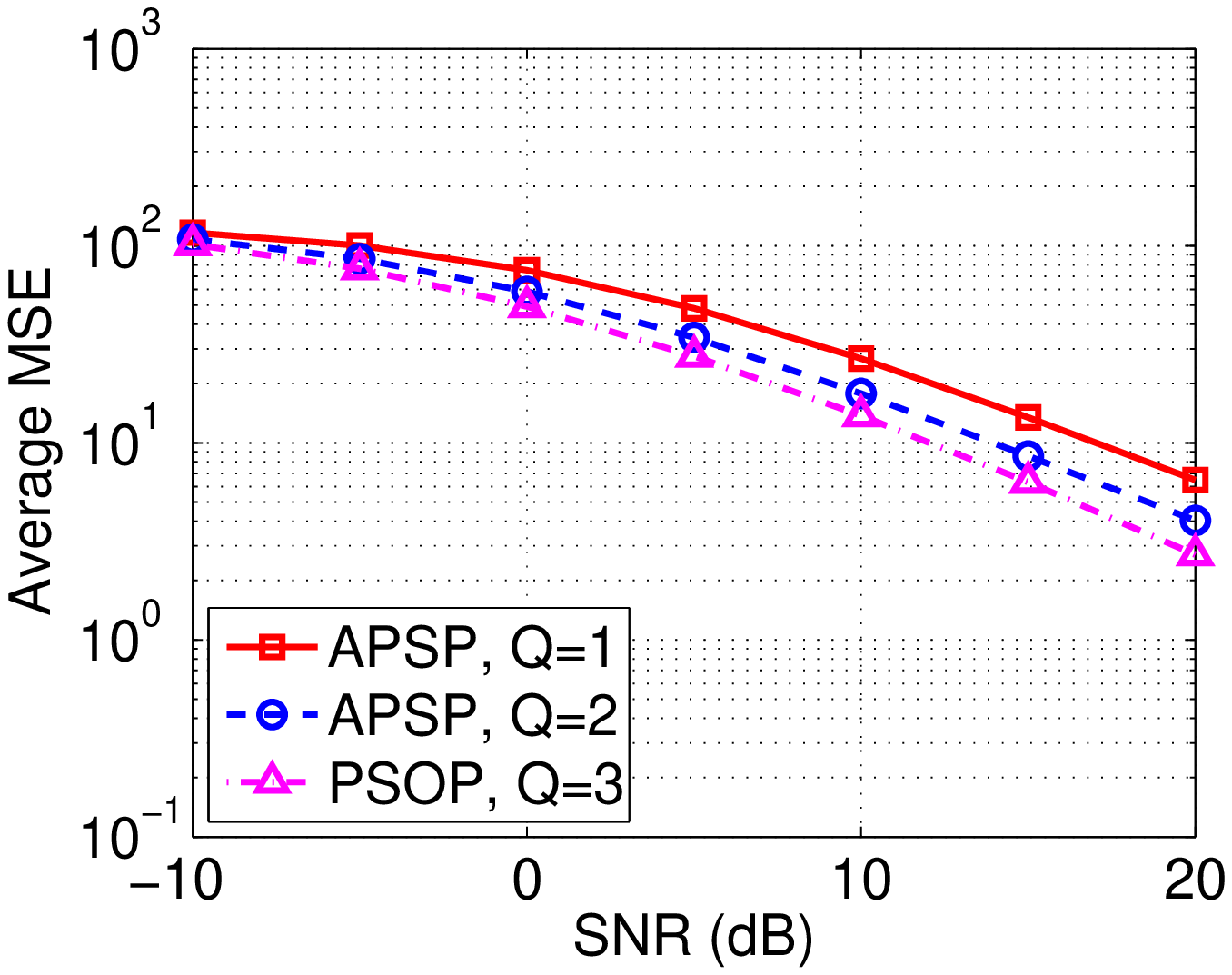}
\label{fig:psmsecomc}}
\caption{Comparison of the pilot segment MSE-CE performance of APSPs and PSOPs. Results are shown versus the pilot SNR in several typical scenarios.}
\label{fig:psmsecom}
\end{figure*}

With the above settings, we compare the performance of the proposed APSP-CA approach with that of the conventional PSOP-CA approach, which serves as the benchmark for comparison of channel acquisition performance. For the conventional PSOP-CA, the required pilot segment length is $Q=\left\lceil K/\left(\dnnot{N}{c}/\dnnot{N}{g}\right)\right\rceil=3$ OFDM symbols \cite{Marzetta10Noncooperative}. For the proposed APSP-CA, the pilot segment length can be set to $Q=1$ or $2$. We adopt \alref{alg:GPPSSA} to schedule the pilot phase shifts in the simulations, and the overlap threshold in the algorithm is set as $\gamma=10^{-4}$. Although this algorithm is suboptimal in general compared with exhaustive search, substantial performance gains over the conventional PSOP-CA in terms of achievable spectral efficiency can still be achieved with relatively little computational cost.

In \figref{fig:psmsecom}, the pilot segment MSE-CE performance\footnote{All the simulated MSE results are normalized by the number of subcarriers $\dnnot{N}{c}$ and the number of UTs $K$.} obtained by the proposed APSPs (with $Q=1$ and $2$) are compared with those for conventional PSOPs ($Q=3$) under several typical propagation scenarios. It can be observed that, in all the considered scenarios, the MSE-CE performance with APSPs approaches the performance obtained with PSOPs, while the pilot overhead is reduced by 66.7\% ($Q=1$) and 33.3\% ($Q=2$), respectively.

\begin{figure*}[!t]
\centering
\subfloat[SU]{\includegraphics[width=0.32\textwidth]{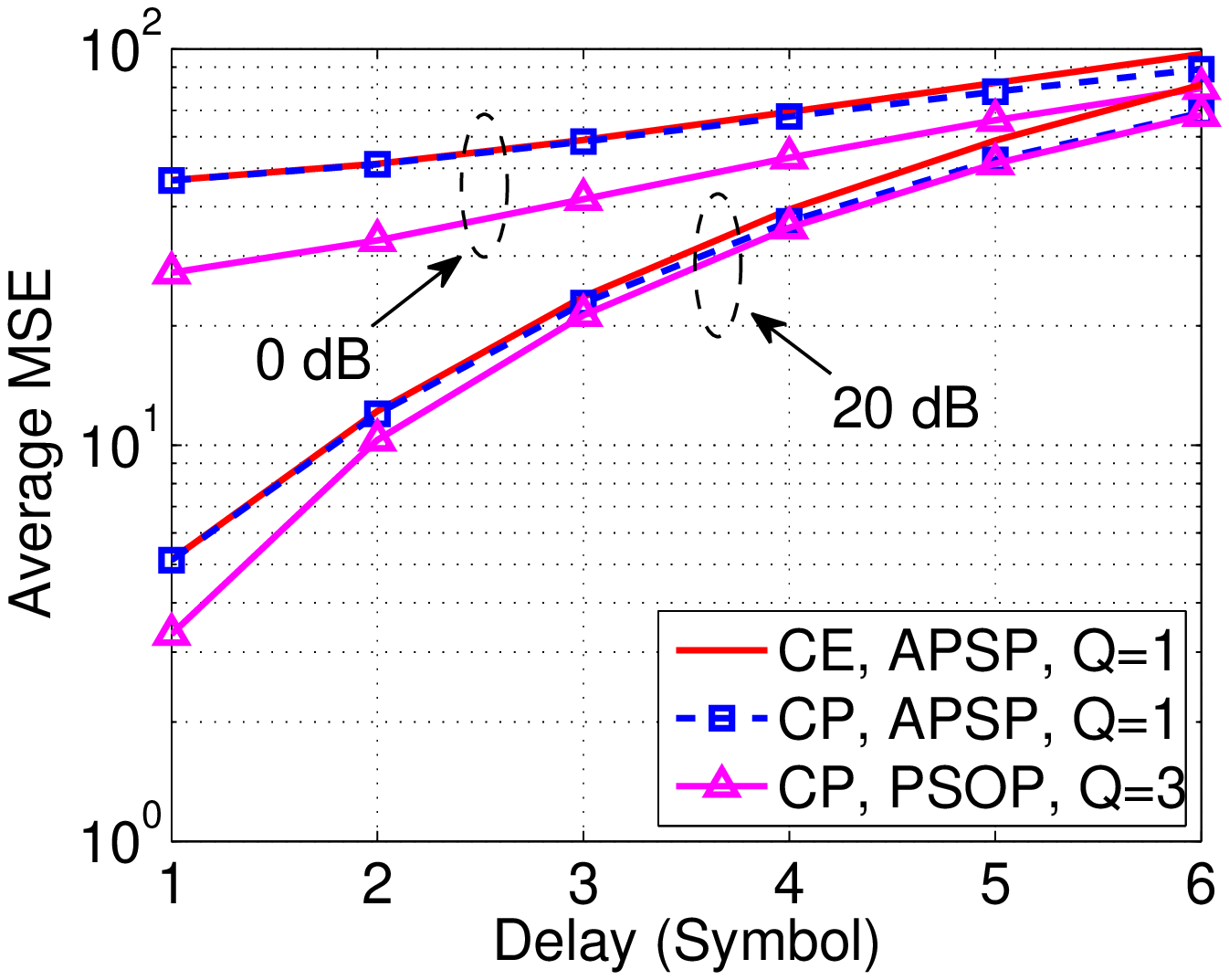}
\label{fig:msecpcompa}}
\hfill
\subfloat[UMa]{\includegraphics[width=0.32\textwidth]{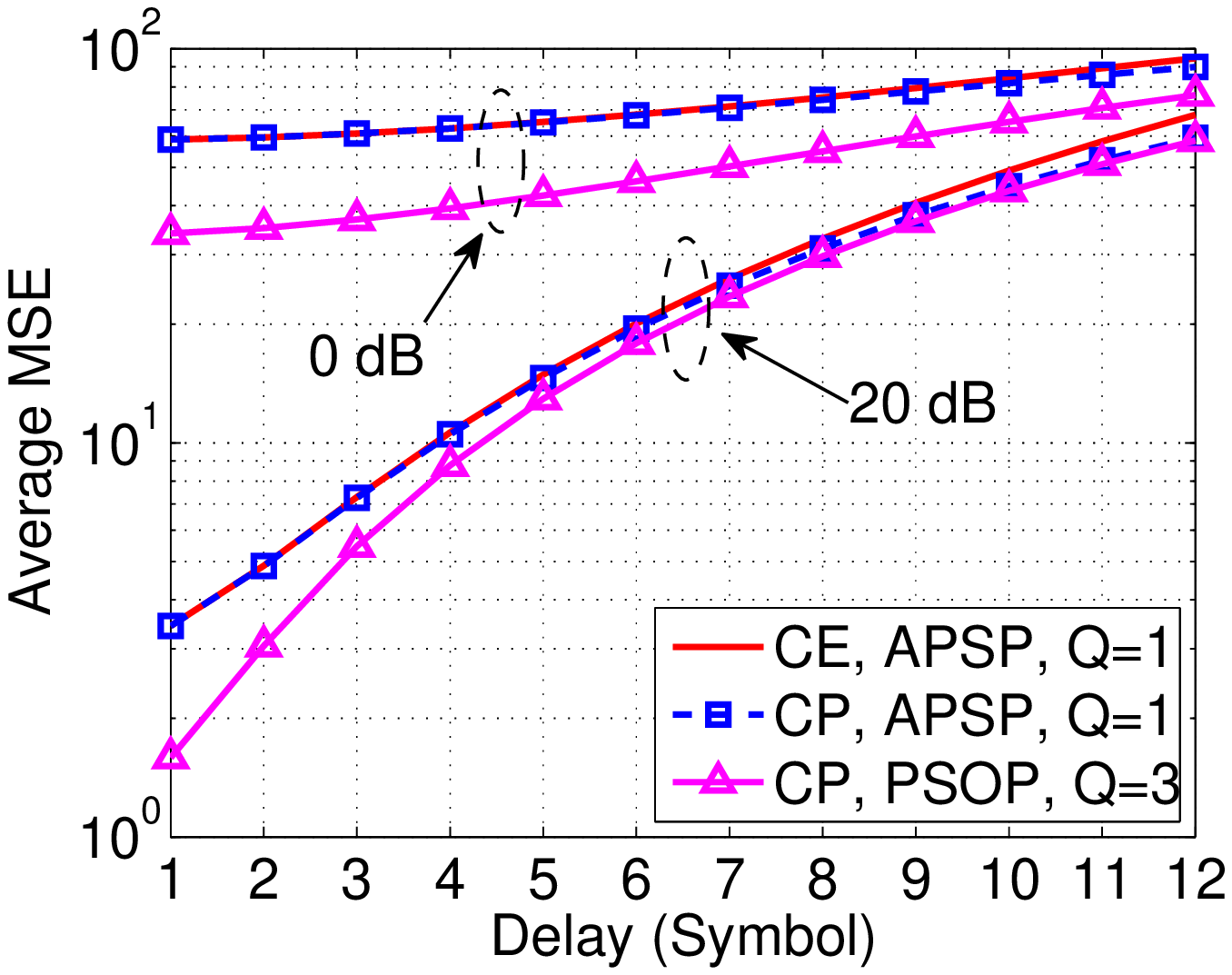}
\label{fig:msecpcompb}}
\hfill
\subfloat[UMi]{\includegraphics[width=0.32\textwidth]{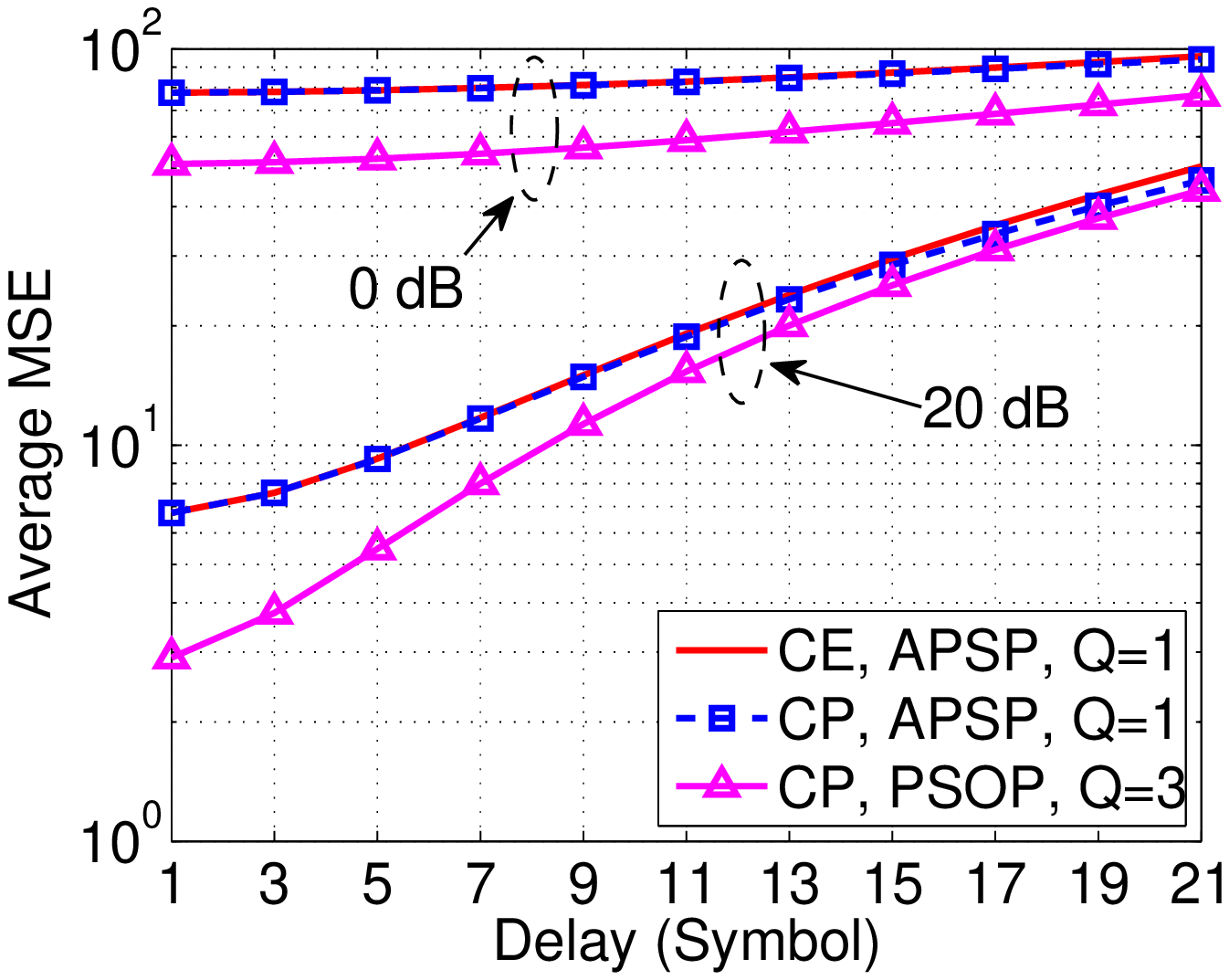}
\label{fig:msecpcompc}}
\caption{Comparison of the channel acquisition performance during the data segment between APSPs and PSOPs. For APSPs, the performance of both channel estimation and prediction are depicted. Results are shown versus the channel acquisition delay in several typical scenarios.}
\label{fig:msecpcomp}
\end{figure*}

In \figref{fig:msecpcomp}, we compare the channel acquisition performance during the data segment in terms of MSE versus the delay $\Delta_{\ell}$ between the data symbol and pilot segment. Both the APSP-CA ($Q=1$) and PSOP-CA ($Q=3$) are evaluated. Also, for APSPs, both the channel estimation and prediction MSE performance are calculated. It can be observed that the MSE-CP performance obtained with APSPs approaches that for PSOPs, with the pilot overhead reduced by 66.7\%. In addition, with APSPs, channel prediction outperforms channel estimation in all the evaluated scenarios. Note that the channel acquisition performance in terms of both MSE-CE and MSE-CP grows almost linearly with delay, and thus the channel acquisition performance can be improved when combined with the type-B frame structure, as shown in the following simulation results.

\begin{figure*}[!t]
\centering
\subfloat[SU]{\includegraphics[width=0.32\textwidth]{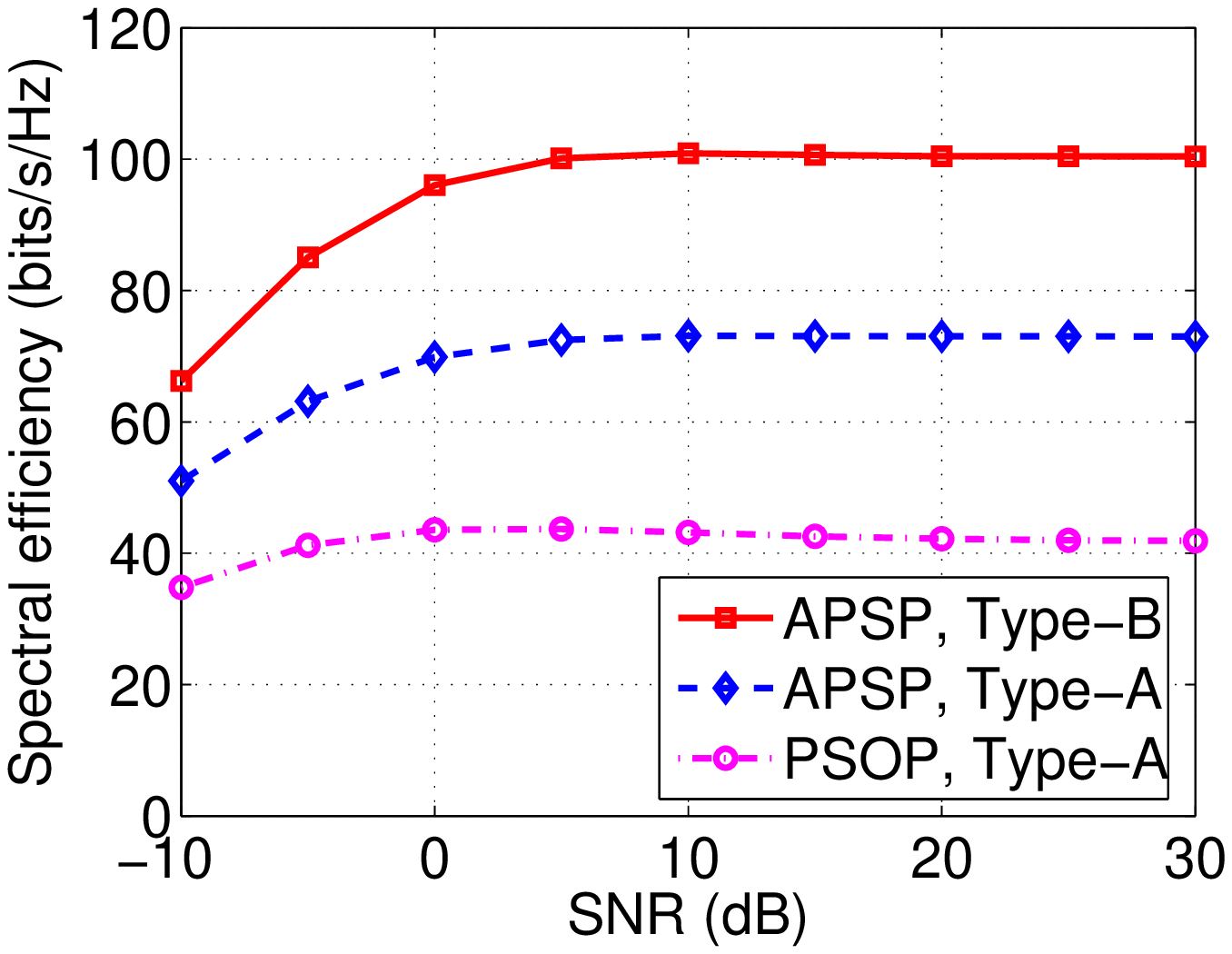}
\label{fig:ratecompa}}
\hfill
\subfloat[UMa]{\includegraphics[width=0.32\textwidth]{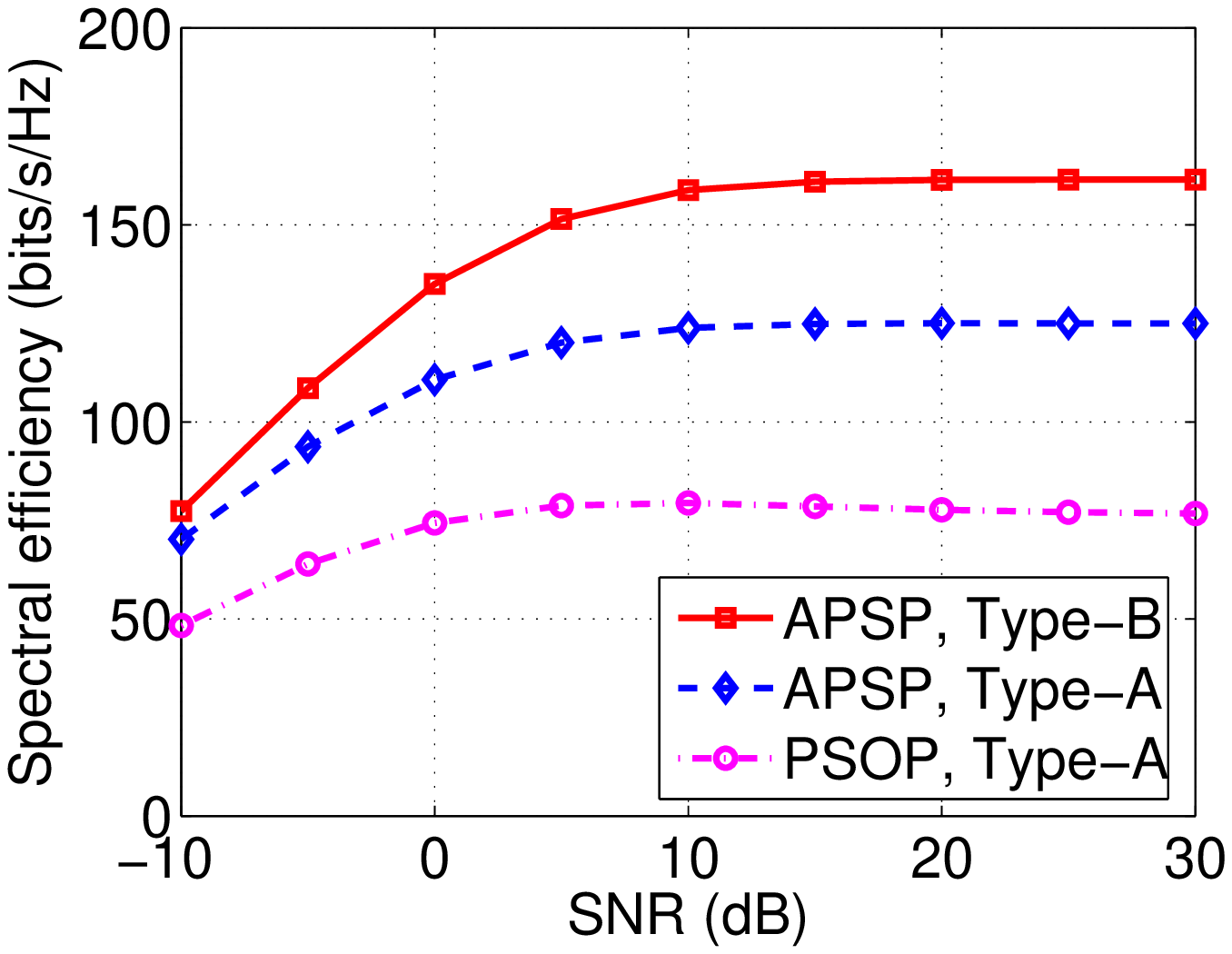}
\label{fig:ratecompb}}
\hfill
\subfloat[UMi]{\includegraphics[width=0.32\textwidth]{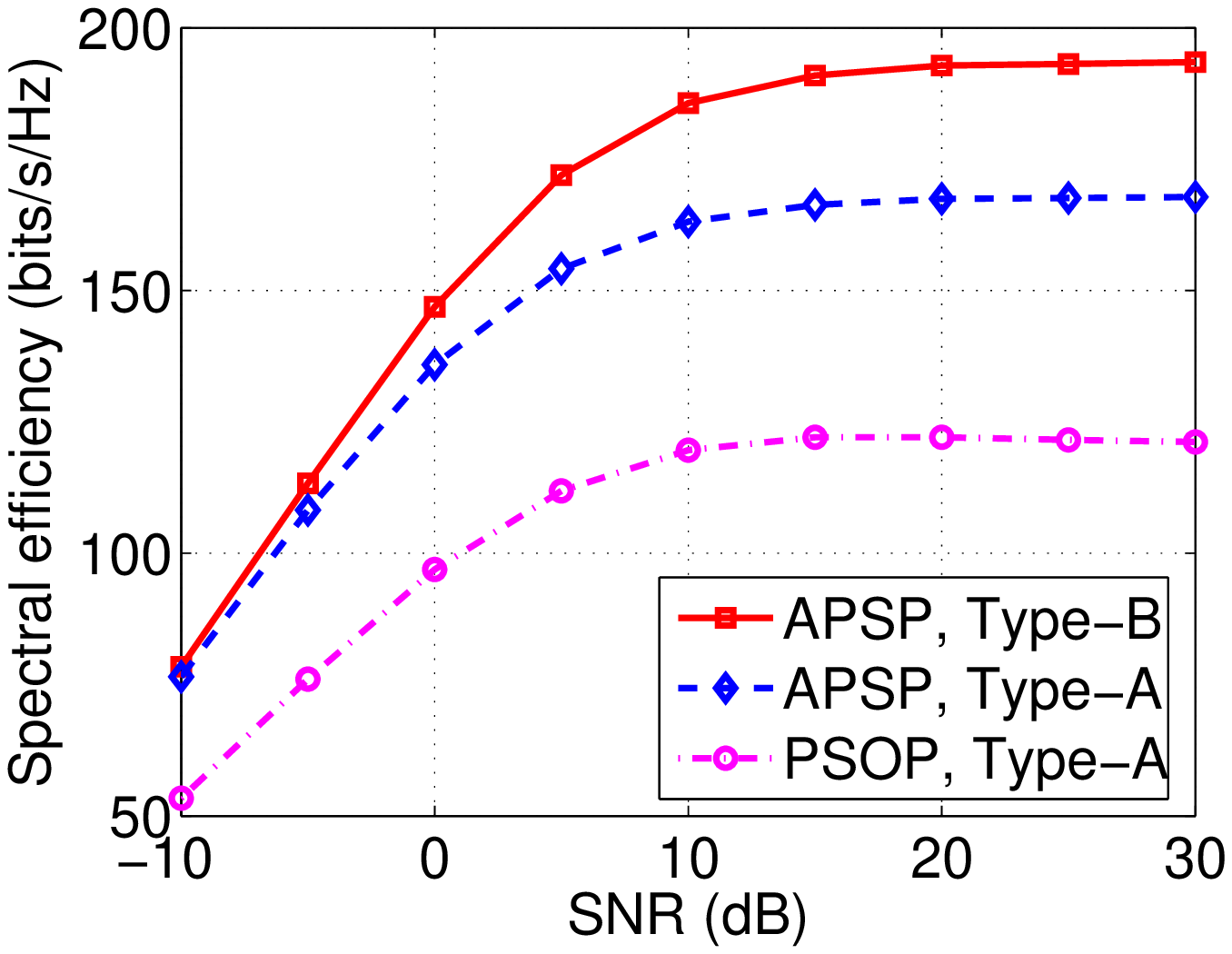}
\label{fig:ratecompc}}
\caption{Comparison of the achievable spectral efficiency between the APSP-CA and PSOP-CA approaches. For the APSP-CA approach, both type-A and type-B frame structures are considered. Results are shown versus the SNR in several typical scenarios.}
\label{fig:ratecomp}
\end{figure*}

At the end of this section, we compare the achievable spectral efficiency of the proposed APSP and the conventional PSOP approaches.\footnote{Note that the achievable spectral efficiency can reflect the tradeoff between the transmission performance and pilot overhead. Intuitively, reducing the pilot overhead decreases the channel acquisition quality (which leads to degradation of the achievable spectral efficiency), but also increases the length of the data segments (which leads to increased achievable spectral efficiency).} We assume that the frame length equals 500 $\mu$s as in \cite{Marzetta10Noncooperative}, which is equal to the length of 7 OFDM symbols \cite{3gpp.36.211}, and that UL and DL data transmission each occupies half of the data segment length. For the conventional PSOP-CA approach, channel estimation and the type-A frame structure in \figref{fig:frame}\subref{fig:conframe} are adopted. For the proposed APSP-CA approach, both APSPs ($Q=1$) and channel prediction are adopted, and both type-A and type-B frame structures are considered. A MMSE receiver and precoder are employed for both UL and DL data transmissions, and the SNR is assumed to be equal to the pilot SNR. In \figref{fig:ratecomp}, the achieved spectral efficiency\footnote{The achievable UL rate is evaluated using the classical worst case approach as in \cite{Hassibi03How}, and the achievable DL rate is evaluated using the approach in \cite{Jose11Pilot}. The OFDM guard interval overhead is taken into account.} of the APSP-CA and PSOP-CA approaches are depicted. It can be observed that the proposed APSP-CA approach shows substantial performance gain in terms of the achievable spectral efficiency over the conventional PSOP-CA approach, especially in the high mobility regime where pilot overhead dominates and the high SNR regime where pilot interference dominates. Specifically, in the high mobility SU scenario (250 km/h) with an SNR of 10 dB, the proposed APSPs can provide about 69\% in average spectral efficiency gains over the conventional PSOPs. In addition, the type-B frame structure can provide a gain of about 64\% over the type-A frame structure when APSPs are adopted.

\section{Conclusion}\label{sec:conc_pw}

In this paper, we proposed a channel acquisition approach with adjustable phase shift pilots (APSPs) for massive MIMO-OFDM to reduce the pilot overhead. We first investigated the channel sparsity in massive MIMO-OFDM based on a physically motivated channel model. With this channel model, we investigated channel estimation and prediction for massive MIMO-OFDM with APSPs, and provided an optimal pilot phase shift scheduling condition applicable to both channel estimation and prediction. We further developed a simplified pilot phase shift scheduling algorithm based on this optimal channel acquisition condition with APSPs. The proposed APSP-CA implemented over both one and multiple symbols were investigated. Significant performance gains in terms of achievable spectral efficiency were observed for the proposed APSP-CA approach over the conventional PSOP-CA approach in several typical mobility scenarios.

\appendices

\section{Derivation of {\eqref{eq:expbgkldel}}}\label{app:der_sta}

The derivation of \eqref{eq:expbgkldel} is detailed in \eqref{eq:derofsta}, shown at the top of the next page, where (a) follows from \eqref{eq:uncorcha}, and (b) follows from the definition of the delta function.
\begin{figure*}
\begin{align}\label{eq:derofsta}
  &\expect{\vecop{\bG_{k,\ell+\Delta_{\ell}}}\vecherop{\bG_{k,\ell}}} \ntb
  &\quad=\mathsf{E}\left\{\left[\sum_{q=0}^{\dnnot{N}{g}-1}\int\limits_{-\infty}^{\infty}\int\limits_{-\pi/2}^{\pi/2}\!
  \left[\bbf_{\dnnot{N}{c},q}\otimes\vtheta\right]\cdot\expx{\barjmath2\pi \nu\left(\ell+\Delta_{\ell}\right)\dnnot{T}{sym}}
  \cdot g_{k}\left(\theta,q\dnnot{T}{s},\nu\right)\intdx{\theta}\intdx{\nu}\right]\right.\ntb
  &\qquad\quad\cdot\left.\left[\sum_{q'=0}^{\dnnot{N}{g}-1}\int\limits_{-\infty}^{\infty}\int\limits_{-\pi/2}^{\pi/2}\!
  \left[\bbf_{\dnnot{N}{c},q'}\otimes\bv_{M,\theta'}\right]^{H}\cdot\expx{-\barjmath2\pi \nu'\ell\dnnot{T}{sym}}
  \cdot g_{k}\left(\theta',q'\dnnot{T}{s},\nu'\right)\intdx{\theta'}\intdx{\nu'}\right]\right\}\ntb
  &\quad=\sum_{q=0}^{\dnnot{N}{g}-1}\sum_{q'=0}^{\dnnot{N}{g}-1}\int\limits_{-\infty}^{\infty}\int\limits_{-\pi/2}^{\pi/2}\int\limits_{-\infty}^{\infty}\int\limits_{-\pi/2}^{\pi/2}\!
  \left[\bbf_{\dnnot{N}{c},q}\otimes\vtheta\right]\left[\bbf_{\dnnot{N}{c},q'}\otimes\bv_{M,\theta'}\right]^{H}
  \cdot\expx{\barjmath2\pi \nu\left(\ell+\Delta_{\ell}\right)\dnnot{T}{sym}}
  \cdot\expx{-\barjmath2\pi \nu'\ell\dnnot{T}{sym}}
  \ntb
  &\qquad
  \cdot \expect{g_{k}\left(\theta,q\dnnot{T}{s},\nu\right)
  g_{k}\left(\theta',q'\dnnot{T}{s},\nu'\right)}\intdx{\theta}\intdx{\nu}\intdx{\theta'}\intdx{\nu'}\ntb
  &\quad\equaa\sum_{q=0}^{\dnnot{N}{g}-1}\sum_{q'=0}^{\dnnot{N}{g}-1}\int\limits_{-\infty}^{\infty}\int\limits_{-\pi/2}^{\pi/2}\int\limits_{-\infty}^{\infty}\int\limits_{-\pi/2}^{\pi/2}\!
  \left[\bbf_{\dnnot{N}{c},q}\otimes\vtheta\right]\left[\bbf_{\dnnot{N}{c},q'}\otimes\bv_{M,\theta'}\right]^{H}
  \cdot\expx{\barjmath2\pi \nu\left(\ell+\Delta_{\ell}\right)\dnnot{T}{sym}}
  \cdot\expx{-\barjmath2\pi \nu'\ell\dnnot{T}{sym}}
  \ntb
  &\qquad
  \cdot \upnot{\ttS_{k}}{AD}\left(\theta,q\dnnot{T}{s}\right)\cdot\upnot{\ttS_{k}}{Dop}\left(\nu\right)\cdot\delfunc{\theta-\theta'}\delfunc{q-q'}\delfunc{\nu-\nu'}\intdx{\theta}\intdx{\nu}\intdx{\theta'}\intdx{\nu'}\ntb
  &\quad\equab\sum_{q=0}^{\dnnot{N}{g}-1}\int\limits_{-\infty}^{\infty}\int\limits_{-\pi/2}^{\pi/2}\!
  \left[\bbf_{\dnnot{N}{c},q}\otimes\vtheta\right]\left[\bbf_{\dnnot{N}{c},q}\otimes\bv_{M,\theta}\right]^{H}
  \cdot\expx{\barjmath2\pi\nu\Delta_{\ell}\dnnot{T}{sym}}
  \cdot \upnot{\ttS_{k}}{AD}\left(\theta,q\dnnot{T}{s}\right)\cdot\upnot{\ttS_{k}}{Dop}\left(\nu\right)\intdx{\theta}\intdx{\nu}\ntb
  &\quad=\underbrace{\int\limits_{-\infty}^{\infty}\!\expx{\barjmath2\pi\nu\Delta_{\ell}\dnnot{T}{sym}} \cdot\upnot{\ttS_{k}}{Dop}\left(\nu\right)\intdx{\nu}}_{\varrho_{k}\left(\Delta_{\ell}\right)}
  \cdot
  \underbrace{\sum_{q=0}^{\dnnot{N}{g}-1}\int\limits_{-\pi/2}^{\pi/2}\!
  \left[\bbf_{\dnnot{N}{c},q}\otimes\vtheta\right]\left[\bbf_{\dnnot{N}{c},q}\otimes\bv_{M,\theta}\right]^{H}
  \cdot\upnot{\ttS_{k}}{AD}\left(\theta,q\dnnot{T}{s}\right)\intdx{\theta}}_{\bR_{k}}
\end{align}
\hrule
\end{figure*}

\section{Proof of {\propref{prop:Decomp_cov}}}\label{app:prop_Decomp_cov}

We start by defining some auxiliary variables to simplify the derivations. We define $n_{d}\triangleq\left\lfloor d/M\right\rfloor$ and $m_{d}\triangleq\modulo{d}{M}$ for an arbitrary non-negative integer $d$. Note that the element indices start from $0$ in this paper. Then we can readily obtain that for a matrix $\bOmega_{k}\in\bbR^{M\times\dnnot{N}{g}}$, the $d$th element of $\vecop{\bOmega_{k}}$ equals the $\left(m_{d},n_{d}\right)$th element of $\bOmega_{k}$, i.e., $\vecele{\vecop{\bOmega_{k}}}{d}=\vecele{\bOmega_{k}}{m_{d},n_{d}}$. We can also obtain that for matrices $\bF\in\bbC^{\dnnot{N}{c}\times\dnnot{N}{g}}$ and $\bV\in\bbC^{M\times M}$, $\vecele{\bF\otimes\bV}{i,j}=\vecele{\bF}{n_i,n_j}\vecele{\bV}{m_i,m_j}$ from the definition of the Kronecker product. With the above definitions and related properties, the proof can be obtained as follows:

\begin{align}\label{eq:pradlm}
  &\lim_{\substack{\toinf{M}}}
  \Big[\bR_{k}-\left(\bF_{\dnnot{N}{c}\times\dnnot{N}{g}}\otimes\bV_{M}\right)
  \diag{\vecop{\bOmega_{k}}}\ntb
  &\qquad\cdot\left(\bF_{\dnnot{N}{c}\times\dnnot{N}{g}}\otimes\bV_{M}\right)^{H}\Big]_{i,j}\ntb
  &\quad=\lim_{\substack{\toinf{M}}}
  \vecele{\bR_{k}}{i,j}-\lim_{\substack{\toinf{M}}}\sum_{d=0}^{M\dnnot{N}{g}-1}\vecele{\vecop{\bOmega_{k}}}{d}\ntb
  &\qquad\cdot\vecele{\bF_{\dnnot{N}{c}\times\dnnot{N}{g}}\otimes\bV_{M}}{i,d}\vecele{\bF_{\dnnot{N}{c}\times\dnnot{N}{g}}\otimes\bV_{M}}{j,d}^{*}\ntb%
  &\quad\equaa\lim_{\substack{\toinf{M}}}\vecele{\bR_{k}}{i,j}-\lim_{\substack{\toinf{M}}}
  \sum_{n_{d}=0}^{\dnnot{N}{g}-1}\sum_{m_{d}=0}^{M-1}\vecele{\bOmega_{k}}{m_{d},n_{d}}\ntb
  &\qquad
  \cdot\vecele{\bF_{\dnnot{N}{c}\times\dnnot{N}{g}}}{n_{i},n_{d}}\vecele{\bF_{\dnnot{N}{c}\times\dnnot{N}{g}}}{n_{j},n_{d}}^{*}
  \vecele{\bV_{M}}{m_{i},m_{d}}\vecele{\bV_{M}}{m_{j},m_{d}}^{*}\ntb%
  &\quad\equab\lim_{\substack{\toinf{M}}}\vecele{\bR_{k}}{i,j}
  -\lim_{\substack{\toinf{M}}}\oneon{M\dnnot{N}{c}}\sum_{n_{d}=0}^{\dnnot{N}{g}-1}\sum_{m_{d}=0}^{M-1}
  M\dnnot{N}{c}\ntb
  &\qquad\cdot\left(\theta_{m_{d}+1}-\theta_{m_{d}}\right)
  \upnot{\ttS_{k}}{AD}\left(\theta_{m_{d}},\tau_{n_{d}}\right)\ntb
  &\qquad\cdot\expx{-\barjmath2\pi\frac{\left(n_{i}-n_{j}\right)n_{d}}{\dnnot{N}{c}}}\ntb
  &\qquad\cdot\expx{-\barjmath2\pi\frac{\left(m_{i}-m_{j}\right)\left(m_{d}-M/2\right)}{M}}\ntb%
  &\quad\equac\sum_{q=0}^{\dnnot{N}{g}-1}
  \int\limits_{-\pi/2}^{\pi/2}\!\left[\bbf_{\dnnot{N}{c},q}\otimes\vtheta\right]_{i}
  \left[\bbf_{\dnnot{N}{c},q}\left(q\right)\otimes\vtheta\right]_{j}^{*}\ntb
  &\qquad\cdot\upnot{\ttS_{k}}{AD}\left(\theta,q\dnnot{T}{s}\right)\intdx{\theta}
  -\lim_{\substack{\toinf{M}}}\sum_{n_{d}=0}^{\dnnot{N}{g}-1}\sum_{m_{d}=0}^{M-1}
  \left(\theta_{m_{d}+1}-\theta_{m_{d}}\right)\ntb
  &\qquad\cdot\upnot{\ttS_{k}}{AD}\left(\theta_{m_{d}},\tau_{n_{d}}\right)
  \expx{-\barjmath2\pi\frac{\left(n_{i}-n_{j}\right)n_{d}}{\dnnot{N}{c}}}\ntb
  &\qquad\cdot
  \expx{-\barjmath\pi\left(m_{i}-m_{j}\right)\sin\left(\theta_{m_{d}}\right)}\ntb
  &\quad\equad\sum_{q=0}^{\dnnot{N}{g}-1}\int\limits_{-\pi/2}^{\pi/2}\!
  \vecele{\bbf_{\dnnot{N}{c},q}}{n_{i}}\vecele{\bbf_{\dnnot{N}{c},q}}{n_{j}}^{*}
  \vecele{\vtheta}{m_{i}}\vecele{\vtheta}{m_{j}}^{*}\ntb
  &\qquad\cdot\upnot{\ttS_{k}}{AD}\left(\theta,\tau_{q}\right)\intdx{\theta}
  -\sum_{r=0}^{\dnnot{N}{g}-1}\int\limits_{\theta_{0}}^{\theta_{M}}\!
  \expx{-\barjmath2\pi\frac{\left(n_{i}-n_{j}\right)}{\dnnot{N}{c}}r}\ntb
  &\qquad\cdot
  \expx{-\barjmath\pi\left(m_{i}-m_{j}\right)\sin\left(\theta\right)}\cdot
  \upnot{\ttS_{k}}{AD}\left(\theta,\tau_{r}\right)\intdx{\theta}\ntb
  &\quad\equae\sum_{q=0}^{\dnnot{N}{g}-1}\int\limits_{-\pi/2}^{\pi/2}\!
  \expx{-\barjmath2\pi\frac{\left(n_{i}-n_{j}\right)}{\dnnot{N}{c}}q}\ntb
  &\qquad\cdot
  \expx{-\barjmath\pi\left(m_{i}-m_{j}\right)\sin\left(\theta\right)}\cdot
  \upnot{\ttS_{k}}{AD}\left(\theta,\tau_{q}\right)\intdx{\theta}\ntb
  &\qquad-\sum_{r=0}^{\dnnot{N}{g}-1}\int\limits_{-\pi/2}^{\pi/2}\!
  \expx{-\barjmath2\pi\frac{\left(n_{i}-n_{j}\right)}{\dnnot{N}{c}}r}\ntb
  &\qquad\cdot
  \expx{-\barjmath\pi\left(m_{i}-m_{j}\right)\sin\left(\theta\right)}\cdot
  \upnot{\ttS_{k}}{AD}\left(\theta,\tau_{r}\right)\intdx{\theta}\ntb
  &\quad=0
\end{align}
where (a) follows from the definition of Kronecker product and the definitions of $m_{d}$ and $n_{d}$, (b) follows from \eqref{eq:cov_eigen_ele} and the definitions of $\bF_{\dnnot{N}{c}\times\dnnot{N}{g}}$ and $\bV_{M}$, (c) follows from \eqref{eq:spa_fre_cov} and the definitions of $\tau_{n}$ and $\theta_{m}$, (d) follows from the definition of the Kronecker product, and (e) follows from \eqref{eq:ula steer vec} and the fact that $\theta_{0}=-\pi/2$ and $\theta_{M}=\pi/2$.

Before concluding the proof, we also have to show that both of the limits in the first equation of \eqref{eq:pradlm} exist and are finite. For this purpose, as can be seen from (e) of \eqref{eq:pradlm}, we only need to show that
\begin{align}
  &\Bigg|\sum_{q=0}^{\dnnot{N}{g}-1}\int\limits_{-\pi/2}^{\pi/2}\!
  \expx{-\barjmath2\pi\frac{\left(n_{i}-n_{j}\right)}{\dnnot{N}{c}}q}\ntb
  &\qquad\cdot\expx{-\barjmath\pi\left(m_{i}-m_{j}\right)\sin\left(\theta\right)}\cdot
  \upnot{\ttS_{k}}{AD}\left(\theta,\tau_{q}\right)\intdx{\theta}\Bigg| \ntb
  &\quad\mathop{\leq}^{(\mathrm{a})}\sum_{q=0}^{\dnnot{N}{g}-1}\Bigg|\int\limits_{-\pi/2}^{\pi/2}\!
  \expx{-\barjmath2\pi\frac{\left(n_{i}-n_{j}\right)}{\dnnot{N}{c}}q}\ntb
  &\quad\qquad\cdot\expx{-\barjmath\pi\left(m_{i}-m_{j}\right)\sin\left(\theta\right)}\cdot
  \upnot{\ttS_{k}}{AD}\left(\theta,\tau_{q}\right)\intdx{\theta}\Bigg| \ntb
  &\quad\mathop{\leq}^{(\mathrm{b})}\sum_{q=0}^{\dnnot{N}{g}-1}\int\limits_{-\pi/2}^{\pi/2}\!
  \Bigg|\expx{-\barjmath2\pi\frac{\left(n_{i}-n_{j}\right)}{\dnnot{N}{c}}q}\ntb
  &\quad\qquad\cdot\expx{-\barjmath\pi\left(m_{i}-m_{j}\right)\sin\left(\theta\right)}\cdot
  \upnot{\ttS_{k}}{AD}\left(\theta,\tau_{q}\right)\Bigg|\intdx{\theta} \ntb
  &\quad=\sum_{q=0}^{\dnnot{N}{g}-1}\int\limits_{-\pi/2}^{\pi/2}\!
  \abs{\upnot{\ttS_{k}}{AD}\left(\theta,\tau_{q}\right)}\intdx{\theta}\ntb
  &\quad\mathop{<}^{(\mathrm{c})}\infty
\end{align}
where (a) follows from the triangle inequality $\abs{\sum_{q=0}^{N-1}a_{q}}\leq\sum_{q=0}^{N-1}\abs{a_{q}}$, (b) follows from the integral property $\abs{\int_{a}^{b}\!f\left(x\right)\intdx{x}}\leq\int_{a}^{b}\!\abs{f\left(x\right)}\intdx{x}$, and (c) follows from the fact that the power angle-delay spectrum function $\upnot{\ttS_{k}}{AD}\left(\theta,\tau\right)$, which represents the channel power in the angle-delay domain, is bounded. This concludes the proof.

\section{Proof of {\propref{prop:adcmsts}}}\label{app:prop_adcmsts}
To show \eqref{eq:exbhk}, it suffices to show that
\begin{align}
  &\expect{\vecop{\bH_{k,\ell+\Delta_{\ell}}}\vecherop{\bH_{k,\ell}}}\ntb
  &\qquad=\varrho_{k}\left(\Delta_{\ell}\right)\cdot\diag{\vecop{\bOmega_{k}}}.
\end{align}
From the definition of $\bH_{k,\ell}$ given in \eqref{eq:adcr}, we can obtain
\begin{align}
  \vecop{\bH_{k,\ell}}&=\left(\bF_{\dnnot{N}{c}\times\dnnot{N}{g}}^{H}\otimes\bV_{M}^{H}\right)\vecop{\bG_{k,\ell}}\ntb
  &=\left(\bF_{\dnnot{N}{c}\times\dnnot{N}{g}}\otimes\bV_{M}\right)^{H}\vecop{\bG_{k,\ell}}
\end{align}
via employing the Kronecker product identities $\vecop{\bA\bB\bC}=\left(\bC^{T}\otimes\bA\right)\vecop{\bB}$ and $\bA^{H}\otimes\bB^{H}=\left(\bA\otimes\bB\right)^{H}$ \cite{Seber08Matrix}.

Then it can be shown that
\begin{align}
  &\expect{\vecop{\bH_{k,\ell+\Delta_{\ell}}}\vecherop{\bH_{k,\ell}}}\ntb
  &\qquad\equaa\left(\bF_{\dnnot{N}{c}\times\dnnot{N}{g}}\otimes\bV_{M}\right)^{H}
  \expect{\vecop{\bG_{k,\ell+\Delta_{\ell}}}\vecherop{\bG_{k,\ell}}}\ntb
  &\qquad\quad\cdot\left(\bF_{\dnnot{N}{c}\times\dnnot{N}{g}}\otimes\bV_{M}\right)\ntb
  &\qquad\equab\varrho_{k}\left(\Delta_{\ell}\right)\cdot\left(\bF_{\dnnot{N}{c}\times\dnnot{N}{g}}\otimes\bV_{M}\right)^{H}
  \bR_{k}\left(\bF_{\dnnot{N}{c}\times\dnnot{N}{g}}\otimes\bV_{M}\right)\ntb
  &\qquad\equac\varrho_{k}\left(\Delta_{\ell}\right)\cdot\diag{\vecop{\bOmega_{k}}}
\end{align}
where (a) follows from the fact that $\bF_{\dnnot{N}{c}\times\dnnot{N}{g}}$ and $\bV_{M}$ are both deterministic matrices, (b) follows from \eqref{eq:expbgkldel}, and (c) follows from \propref{prop:Decomp_cov}. This concludes the proof.

\section{Proof of {\propref{prop:mmsetrseqcon}}}\label{app:prop_mmsetrseqcon}

Due to the fact that the elements of $\bOmega_{k'}^{\phi_{k'}-\phi_{k}}$ are non-negative, we can obtain
\begin{align}
  \upnot{{\epsilon}}{CE}=&\sum_{k=0}^{K-1}
  \sum_{i=0}^{M-1}\sum_{j=0}^{\dnnot{N}{g}-1}\Bigg\{\vecele{\bOmega_{k}}{i,j}
  \ntb
  &\qquad-\frac{\vecele{\bOmega_{k}}{i,j}^{2}}{\vecele{\bOmega_{k}}{i,j}+\sum_{k'\neq k}\vecele{\bOmega_{k'}^{\phi_{k'}-\phi_{k}}}{i,j}+\oneon{\dnnot{\rho}{tr}}}\Bigg\}\ntb
  \geq&\sum_{k=0}^{K-1}
  \sum_{i=0}^{M-1}\sum_{j=0}^{\dnnot{N}{g}-1}\left\{\vecele{\bOmega_{k}}{i,j}
  -\frac{\vecele{\bOmega_{k}}{i,j}^{2}}{\vecele{\bOmega_{k}}{i,j}+\oneon{\dnnot{\rho}{tr}}}\right\}
  =\upnot{\varepsilon}{CE}.
\end{align}
Furthermore, when the condition $\left(\barbOmega_{k,\singbrac{\dnnot{N}{c}}}\bPi_{\dnnot{N}{c}}^{\phi_{k}}\right)
\odot\left(\barbOmega_{k',\singbrac{\dnnot{N}{c}}}\bPi_{\dnnot{N}{c}}^{\phi_{k'}}\right)=\bzero$ is satisfied, then with the same column permutation and column truncation, multiplications of the corresponding elements still equal zero, i.e.,
\begin{align}\label{eq:condomeg1}
&\left(\barbOmega_{k,\singbrac{\dnnot{N}{c}}}\bPi_{\dnnot{N}{c}}^{\phi_{k}}
\bPi_{\dnnot{N}{c}}^{-\phi_{k}}\bI_{\dnnot{N}{c}\times\dnnot{N}{g}}\right)\ntb
&\qquad\odot\left(\barbOmega_{k',\singbrac{\dnnot{N}{c}}}\bPi_{\dnnot{N}{c}}^{\phi_{k'}}
\bPi_{\dnnot{N}{c}}^{-\phi_{k}}\bI_{\dnnot{N}{c}\times\dnnot{N}{g}}\right)=\bzero.
\end{align}

Recalling the definition in \eqref{eq:omeuphuk} and exploiting the permutation matrix property that $\bPi_{N}^{a}\bPi_{N}^{b}=\bPi_{N}^{a+b}$, the condition in \eqref{eq:condomeg1} is equivalent to
\begin{equation}
  \bOmega_{k}\odot\bOmega_{k'}^{\phi_{k'}-\phi_{k}}=\bzero.
\end{equation}
Then for $\forall i,j$,
\begin{align}
  &\vecele{\bOmega_{k}}{i,j}^{2}\left\{\vecele{\bOmega_{k}}{i,j}+\sum_{k'\neq k}\vecele{\bOmega_{k'}^{\phi_{k'}-\phi_{k}}}{i,j}+\oneon{\dnnot{\rho}{tr}}\right\}\ntb
  &\qquad=\vecele{\bOmega_{k}}{i,j}^{2}\left\{\vecele{\bOmega_{k}}{i,j}+\oneon{\dnnot{\rho}{tr}}\right\}
\end{align}
which leads to
\begin{equation}\label{eq:okijom}
  \frac{\vecele{\bOmega_{k}}{i,j}^{2}}{\vecele{\bOmega_{k}}{i,j}+\sum_{k'\neq k}\vecele{\bOmega_{k'}^{\phi_{k'}-\phi_{k}}}{i,j}+\oneon{\dnnot{\rho}{tr}}}
  =\frac{\vecele{\bOmega_{k}}{i,j}^{2}}{\vecele{\bOmega_{k}}{i,j}+\oneon{\dnnot{\rho}{tr}}}.
\end{equation}
Substituting \eqref{eq:okijom} into \eqref{eq:epsilonkl}, the MSE-CE expression $\upnot{\epsilon_{k}}{CE}$  reduces to
\begin{align}
  \upnot{\varepsilon_{k}}{CE}
  =\sum_{i=0}^{M-1}\sum_{j=0}^{\dnnot{N}{g}-1}\left\{\vecele{\bOmega_{k}}{i,j}
  -\frac{\vecele{\bOmega_{k}}{i,j}^{2}}{\vecele{\bOmega_{k}}{i,j}+\oneon{\dnnot{\rho}{tr}}}\right\}.
\end{align}
Then the minimum in \eqref{eq:varepsilonl} can be achieved. This concludes the proof.

%


\end{document}